\documentclass[12pt,eps]{iopart}

\usepackage{amssymb}
\usepackage[dvips]{graphicx}

\begin{document}

\newcommand \be  {\begin{equation}}
\newcommand \bea {\begin{eqnarray} \nonumber }
\newcommand \ee  {\end{equation}}
\newcommand \eea {\end{eqnarray}}
\renewcommand{\qbeziermax}{1000}

\title[Multifractality and Freezing]{Multifractality and Freezing Phenomena in
Random Energy Landscapes: an Introduction \footnote{Lectures at
International Summer School {\bf "Fundamental Problems in
Statistical Physics XII"} held on August 31 - September 11, 2009 at
Leuven, Belgium}}

\vskip 0.2cm
\author{Yan V Fyodorov\footnote{e-mail: yan.fyodorov@nottingham.ac.uk}}

 \noindent\small{ School of Mathematical Sciences,
University of Nottingham, Nottingham NG72RD, England}

\begin{abstract}
 We start our lectures with introducing and discussing
the general notion of multifractality spectrum for random measures on lattices, and how it can be probed
using moments of that measure. Then we show that the Boltzmann-Gibbs probability distributions
 generated by logarithmically correlated random potentials provide a simple yet
 nontrivial example of disorder-induced multifractal measures.
The typical values of the multifractality exponents can be extracted from calculating the free energy of
the associated Statistical Mechanics problem. To succeed in such a calculation
 we introduce and discuss in some detail two analytically tractable models for
 logarithmically correlated potentials. The first model uses a special definition of distances between points in space
and is based on the idea of multiplicative cascades which originated
in theory of turbulent motion. It is essentially equivalent to statistical mechanics of directed polymers on
disordered trees studied long ago by B. Derrida and H. Spohn in \cite{DS}. In this way we introduce the notion of
the freezing transition which is identified with an abrupt change
in the multifractality spectrum. Second model which allows for explicit analytical evaluation of the free energy
 is the infinite-dimensional version of the problem which can be solved by employing the replica trick.
In particular, the latter version allows one to identify the freezing phenomenon with a mechanism of the replica
symmetry breaking (RSB) and to elucidate its physical meaning. The corresponding 1-step RSB
solution turns out to be {\it marginally stable} everywhere in the low-temperature phase.
We finish with a short discussion of recent developments and extensions of
models with logarithmic correlations, in particular in the context of extreme value statistics.
The first appendix summarizes the standard elementary information about Gaussian integrals and related subjects, and introduces
the notion of the Gaussian Free Field characterized by logarithmic correlations.  Three other appendices
provide the detailed exposition of a few technical details underlying the replica analysis of the model discussed
in the lectures.

\end{abstract}


\maketitle

\section{Introduction}

Investigations of multifractal measures of diverse origin is for several decades a
very active field of research in various branches of applied mathematical sciences
like chaos theory, geophysics, oceanology, climate studies, and finance, and in such areas of
physics as turbulence and statistical mechanics \cite{PV}, and theory of quantum disordered systems
\cite{EM}. The main characteristics of multifractal patterns of data is to possess high variability over
a wide range of space or time scales, associated with huge fluctuations in intensity which can be visually
detected (see fig. 1). Another common feature is presence of certain long-ranged powerlaw-type
correlations in data values.

\vspace{-2ex}

\begin{figure}[h]
\begin{center}
\includegraphics[ scale=0.40]{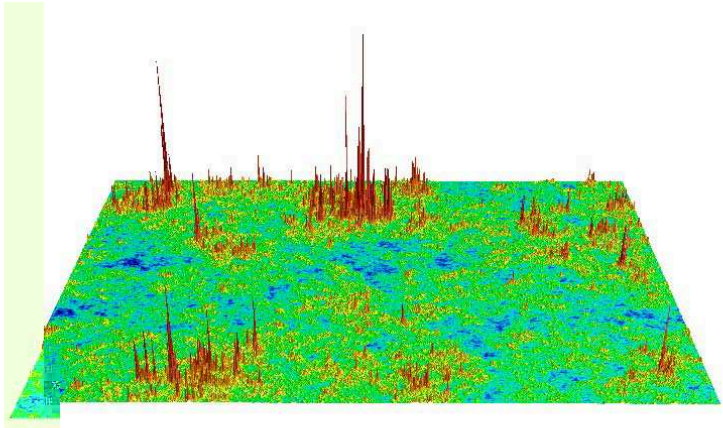}
\caption{\small  Multifractal probability density for a model of quantum particle at the critical point
of Quantum Hall Effect, see \cite{EM}. Courtesy of F. Evers,  A. Mirlin and A. Mildenberger, unpublished.}
\end{center}
\end{figure}

To set the notations, consider a
certain (e.g. hypercubic) lattice of linear extent $L$ in
$N-$dimensional space, with $M\sim L^N$ standing for the total
number of sites in the lattice. The measures of interest are
usually defined via weights $ p_i$ associated with every
lattice site $i=1,2,\ldots, M$ and appropriately normalized to the total weight
equal to unity as sketched below:
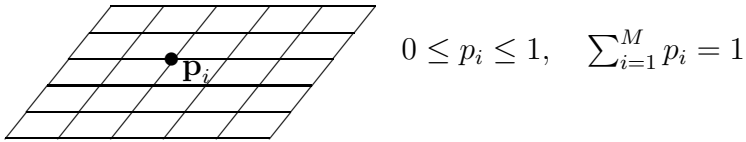
\begin{figure}[h]
\begin{picture}(80,80)(0,-15)
\put(63,30){\circle*{5}}
\put(150,30){$0\le p_i\le 1,\quad \sum_{i=1}^Mp_i=1$}
\put(67,24){${\bf p}_i$}
\multiput(0,0)(20,0){6}{\line(4,5){40}}
\multiput(0,0)(8,10){6}{\line(1,0){100}}
\end{picture}
\caption{A square lattice with weights attributed to the lattice sites.}
\end{figure}

One can imagine a few different spatial arrangements of weights
$p_i$ across the lattice sites. In the case of {\it simply
extended} measures the weights are of similar magnitude at each
lattice site, the normalisation condition then implying the scaling
$p_i\sim M^{-1}$ in the large-$M$ limit. As a generalisation of the
above example one can imagine the non-zero weights $p_i$ supported
evenly on a fractal subset of lattice sites of effective dimension
$0\le N_{ef}<N$. In the limiting case of $N_{ef}=0$ we then deal
with {\it localised} measures characterized by the weights $p_i$
essentially different from zero only inside one or few blobs of
finite total volume. In such a situation weights stay finite even
when $M\to \infty$, that is $p_i=O(M^0)$. Finally, in the most
interesting case of {\it multifractal} measures the weights scale
differently at different sites: $p_i\sim M^{-\alpha_i}$
\footnote{Usually one defines exponents via the relation $p_i\sim
L^{-N\alpha_i}$ i.e. by the reference to linear scale $L$ instead of
the total number of sites $M\sim L^N$. We however find it more
convenient to get rid of trivial spatial dimension factor $N$, and
concentrate only on {\it essential} parameter behaviour.} The full
set of exponents $0\le \alpha_i<\infty$ can be conveniently
characterized by the density $
\rho(\alpha)=\sum_{i=1}^M\,\delta(\alpha-\alpha_i)$ whose scaling
behaviour in the large-$M$ limit is expected to be nontrivial:
$\rho(\alpha)\sim M^{f(\alpha)}$, with the convex function
$f(\alpha)$ known in this context as the {\it multifractality
spectrum} or {\it singularity spectrum}, see Fig. 3. In view of the
identity $\int_{0}^{\infty}\rho(\alpha)\,d\alpha\equiv M$ we see
that at the point of maximum $\alpha=\alpha_0$ we must have
$f(\alpha_0)=1$. Note also that the total number
$m(\alpha)=\int_{0}^{\alpha}\rho(\alpha)\,d\alpha$ of sites of the
lattice characterized by the scaling exponents
$\alpha_i<\alpha(<\alpha_0)$ satisfies for $M\gg 1$  the inequality
$m(\alpha)\sim M^{f(\alpha)}\ge 1$, hence $f(\alpha)\ge 0$ for
$\alpha<\alpha_0$. Modifying this argument one can show
$f(\alpha)\ge 0$ also for $\alpha>\alpha_0$. The condition
$f(\alpha)= 0$ defines generically the minimal $\alpha_{-}$ and
maximal $\alpha_{+}$ threshold values of the exponents which can be
observed in a given typical pattern. Note that the constraint
$p_i\le 1$ implies $\alpha_{-}\ge 0$.

\begin{figure}[h]
\begin{picture}(180,180)(-160,-10)
\put(0,0){\thicklines\line(1,0){170}} \put(0,150){\thicklines \line(1,0){170}}
\put(0,0){\thicklines\line(0,1){150}} \put(170,0){\thicklines \line(0,1){150}}
\put(25,10){\begin{picture}(20,20)
\multiput(00,110)(15,0){6}{\line(1,0){3}}
\put(0,110){\circle*{2}}
\put(-8,106){$1$}
\put(-8,20){$0$}
\put(24,126){$f(\alpha)$}
\put(30,13){$\alpha_{-}$}
\put(100,13){$\alpha_{+}$}
\put(75,13){$\alpha_{0}$}
\put(135,18){$\alpha$}
\multiput(82,20)(0,15){6}{\line(0,1){3}}
\put(0,0){\vector(0,1){130}}
\put(0,20){\vector(1,0){130}}
\qbezier[1000](30,20)(100,200)(100,20)
\end{picture}}
\end{picture}
\caption{ Shape of a typical multifractality spectrum.}
\end{figure}
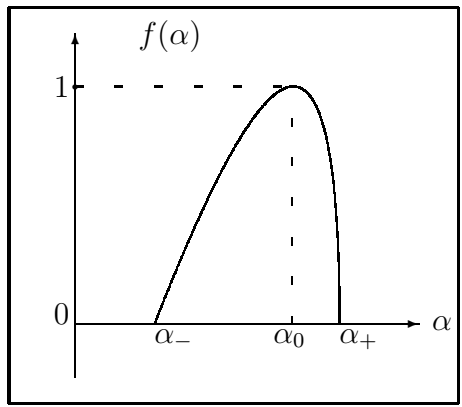

An alternative, frequently more practical way of describing
multifractality  is via the set of exponents $\tau_q$
characterizing the large-$M$ behaviour of the so-called inverse
participation ratios (IPR's) $P_q$ which are simply the moments of
the corresponding measure:
\begin{equation}\label{1}
P_q=\sum_{i=1}^M\, p_i^q=\int\, M^{-q\alpha} \rho(\alpha)\,
d\alpha\,.
\end{equation}
Substituting in the above definition the relation
$\rho(\alpha)\sim M^{f(\alpha)}$ one can evaluate the integral in
the large-$M$ limit by the the steepest descent (also known as Laplace) method, see Appendix A.
One then finds the relation between $\tau_q$ and $f(\alpha)$ given by the {\it \bf Legendre
transform}:
\begin{equation}\label{1a}
P_q\sim M^{-\tau_q},\quad \tau_q=q\alpha-f(\alpha) \quad
\mbox{where}\quad q = \frac{df}{d\alpha}\,\,.
\end{equation}
In particular, at the point of maximum $q=0$ and as from the very definition $\tau_0=-1$ we immediately see that
$f(\alpha_0)\equiv \max_{\alpha}\{f(\alpha)\}=1$, cf. Fig. 3.

The above description is valid for multifractal measures of any
nature. In recent years important insights were obtained for
disorder-generated multifractality, see \cite{EM} and \cite{FRL}
 for a comprehensive discussion in the context of
Anderson localisation transitions, and \cite{MG,F09} for examples
related to Statistical Mechanics in disordered media which are
closer to the context of the present lectures. One of the specific
features of multifractality in the presence of disorder is a
possibility of existence of two different sets of exponents,
$\tau_q$ versus $\tilde{\tau}_q$, governing the scaling behaviour of
typical $P_q$ versus disorder averaged IPR's, $<P_q>\sim
M^{-\tilde{\tau}_q}$. So by definition
\begin{equation}\label{1b}
\tau_q=- \frac{\left\langle\ln{P_q}\right\rangle}{\ln{M}}, \quad
\tilde{\tau_q}=- \frac{\ln{\left\langle P_q \right\rangle}}{\ln{M}}, \quad
\end{equation}
Here and henceforth the brackets stand for the averaging over
different realisations of the disorder.
The first type of averaging featuring in the above equation
 is traditionally called in the literature "quenched" , and
second one is  known as "annealed". It is known that the "quenched" values correspond to
values of exponents which one finds in a "typical" realisation of disorder.
The possibility of "annealed" average to produce results different from typical is related
to a possibility of disorder-averaged moments to be dominated by
exponentially rare configurations in some parameter range. A related aspect of the problem
is that the "annealed" multifractality spectrum
 recovered from the multifractal exponents $\tilde{\tau}_q$ via the Legendre transform
(\ref{1}) can be negative: $\tilde{f}(\alpha)<0$, see fig. 4.
\begin{figure}[h]
\begin{picture}(200,200)(-160,-30)
\put(0,-20){\thicklines\line(1,0){170}} \put(0,150){\thicklines \line(1,0){170}}
\put(0,-20){\thicklines\line(0,1){170}} \put(170,-20){\thicklines \line(0,1){170}}
\put(25,10){\begin{picture}(20,20)
\multiput(00,110)(15,0){6}{\line(1,0){3}}
\put(0,110){\circle*{2}}
\put(-8,106){$1$}
\put(-8,20){$0$}
\put(4,126){$\tilde{f}(\alpha)$}
\put(30,13){$\alpha_{-}$}
\put(100,13){$\alpha_{+}$}
\put(75,13){$\alpha_{0}$}
\put(135,18){$\alpha$}
\multiput(82,20)(0,15){6}{\line(0,1){3}}
\qbezier[1000](30,20)(100,200)(100,20)
\multiput(14,-19)(1,2.5){16}{\circle*{1}}
\multiput(100,20)(0.15,-4.5){8}{\circle*{1}}
\put(0,-10){\vector(0,1){140}}
\put(0,20){\vector(1,0){130}}
\end{picture}}
\end{picture}
\caption{Shape of an "annealed" multifractality spectrum with negative parts (dotted) extracted from the disorder-averaged moments
and reflecting exponentially rare events, see the text.}
\end{figure}
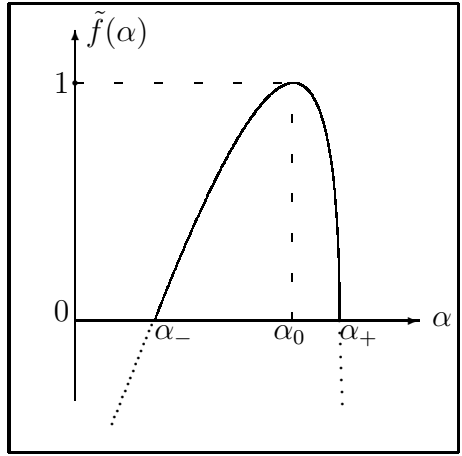

Indeed, those
values reflect events which are exponentially rare \cite{negf} and
need exponentially many realisations of disorder to be observed
experimentally or numerically. On the other hand, as was noted in \cite{EM},
when dealing with {\it typical} multifractality spectrum $f^{typ}(\alpha)$
 by exploiting the relation (\ref{1}) one has to specify the limits of integration over $\alpha$ to be precisely
$\alpha_{-}\le \alpha \le \alpha_+$. IPR moments are then given by
\begin{equation}\label{typ1}
P^{typ}_q=\int_{\alpha_{-}}^{\alpha_{+}}\, M^{-q\alpha+f^{typ}(\alpha)} d\alpha \sim M^{-\tau^{typ}_q}\,,
 \end{equation}
 and calculating the above integral by the steepest descent method reveals that typical (that is {\it quenched})
 exponents $\tau_q=\tau_q^{typ}$
 are related to $f^{typ}(\alpha)$ by Legendre transform only in the range
 $\frac{df}{d\alpha}|_{\alpha_{+}}=q_{min}\le q\le q_{max} =\frac{df}{d\alpha}|_{\alpha_{-}}$, whereas outside that interval
 the exponents behave linearly in $q$, that is $\tau^{typ}_q=q\alpha_{\pm}$, see fig. 5.
We will not dwell on the differences "quenched" vs. "annealed"
exponents further and direct the interested reader to the recent
works \cite{F09} and \cite{FRL} for more detail and further
references \footnote{Note that unfortunately the definitions of the
{\it termination} of the multifractality spectrum used in \cite{F09}
and in \cite{FRL} are essentially different.
 The work \cite{F09} uses the definitions set up in the comprehensive review \cite{EM} which could be consulted in case of confusion.}.
In the present set of lectures we will
concentrate exclusively on calculating  typical (="quenched") values of IPR exponents for some class of models.

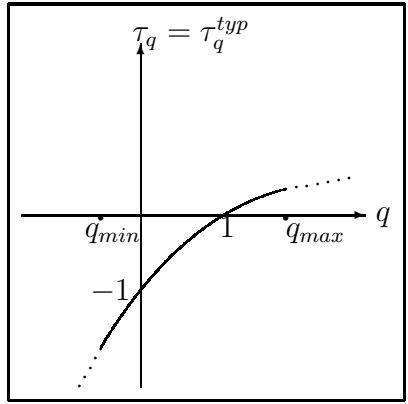
\begin{figure}[h]
\begin{picture}(150,170)(-120,-10)

\put(0,0){\thicklines\line(1,0){150}} \put(0,150){\thicklines \line(1,0){150}}
\put(0,0){\thicklines\line(0,1){150}} \put(150,0){\thicklines \line(0,1){150}}
\put(50,5){\vector(0,1){130}}
\put(5,70){\vector(1,0){130}}

\multiput(105,80)(4,0.7){7}{\circle*{1}}
\multiput(27,5.5)(2,3.5){5}{\circle*{1}}

\put(25,10){\begin{picture}(20,20) \put(10,59){\circle*{1.5}}
\put(80,59){\circle*{1.5}} \put(22,126){$\tau_q=\tau^{typ}_q$}
\put(4,52){$q_{min}$} \put(6,28){$-1$} \put(55,52){$1$}
\put(80,52){$q_{max}$} \put(114,58){$q$}
\qbezier[2000](10,10)(40,60)(80,70)
\end{picture}}
\end{picture}
\caption{$q$-dependence of typical ("quenched") multifractality exponents $\tau_q$. Dotted lines show linear behaviour, see the text.}
\end{figure}

Introduced through the moments involving summation over all the lattice sites, cf. (\ref{1}),
the multifractality by itself says nothing about more delicate questions, for example about spatial correlations between
weights at two different sites of the lattice with coordinates, say, ${\bf x}_1$ and ${\bf x}_2$, separated  by a given distance $|{\bf x}_1-{\bf x}_2|$.
The most natural assumption which is satisfied by vast
majority of multifractal measures of actual experimental interest is the power-law
decay of correlations implied by full statistical {\it spatial self-similarity} of the random measure:
\begin{equation}\label{powerlaw}
\left\langle p^q({\bf x}_1)p^s({\bf x}_2)\right\rangle\propto L^{-N\,y(q,s)}\delta^{-N\, z(q,s)}, \quad \delta=|{\bf x}_1-{\bf x}_2|\,.
\end{equation}
As statistical homogeneity of the random measure implies  for local averages $\left\langle p^q({\bf x}_1)\right\rangle=\frac{1}{M}\sum_{i=1}^M\, p_i^q\sim L^{-N-N\tau_q}$
the equation (\ref{powerlaw}) after setting $\delta\sim L$ yields the relation for exponents:
\begin{equation}\label{powerlaw1}
y(q,s)+z(q,s)-2=\tau_q+\tau_s
\end{equation}
which follows from assuming the decoupling $\left\langle p^q({\bf x}_1)p^s({\bf x}_2)\right\rangle\approx  \left\langle p^q({\bf x}_1)
\right\rangle \left\langle p^s({\bf x}_2)\right\rangle$ at large separations $\delta=|{\bf x}_1-{\bf x}_2|\sim L\to \infty$.
On the other hand, for sites separated by a single lattice spacing $\delta=1$ we must have $\left\langle p^q({\bf x}_1)p^s({\bf x}_2)\right\rangle\approx  \left\langle p^{q+s}({\bf x}_1)
\right\rangle\sim L^{-N-N\tau_{q+s}}$, which after comparing with (\ref{powerlaw}) allows one to relate the exponents governing the spatial correlations to the multifractality exponents as \cite{spatial}
\begin{equation}\label{powerlaw2}
y(q,s)=1+\tau_{q+s},\quad z(q,s)=1+\tau_q+\tau_s-\tau_{q+s}
\end{equation}

Further, it turns out to be instructive to exploit (\ref{powerlaw}) for evaluating the following
correlation function:
\begin{equation}\label{powerlaw3}
\left\langle \ln{p({\bf x}_1)}\ln{p({\bf x}_2)}\right\rangle=\frac{\partial^2}{\partial q\partial s}\left\langle p^q({\bf x}_1)p^s({\bf x}_2)\right\rangle|_{q=s=0}
\end{equation}
Remembering $\tau_0=-1$, and $\left\langle \ln{p({\bf x}_1)}\right\rangle=-N\ln{L}\frac{\partial\tau_q}{\partial q}|_{q=0}$
we obtain after straightforward manipulations the following fundamental relation
\begin{equation}\label{powerlaw4}
\fl \left\langle \ln{p({\bf x}_1)}\ln{p({\bf x}_2)}\right\rangle-\left\langle \ln{p({\bf x}_1)}\right\rangle\left\langle\ln{p({\bf x}_2)}\right\rangle
=-g^2\,\ln{\frac{|x_1-x_2|}{L}}, \quad g^2=N\frac{\partial^2\tau_{q+s}}{\partial q\partial s}|_{q=s=0}>0
\end{equation}
valid for arbitrary self-similar multifractal field. In other words, we have demonstrated that
multifractality plus statistical selfsimilarity and homogeneity of the random weights necessarily imply that the logarithms of such weights
must be correlated {\it logarithmically}  in space.

Inverting such an argument suggests that possibly the simplest way to generate random multifractal weights in the lattice
is by constructing quantities $\ln{p_i}$ at every lattice site $i$ as Gaussian-distributed random variables correlated
in precisely the way prescribed by (\ref{powerlaw4}).
The resulting model has a very natural interpretation in terms of the equilibrium
  statistical mechanics. Indeed, consider a single classical particle subject to a random Gaussian
potential $V({\bf x})$. It is the standard fact of theory of random processes\cite{vanKampen}
that if such a particle moves under the influence of
the thermal white noise according to the Langevin equation
$$\dot{\bf x}=-\frac{\partial}{\partial {\bf x}}V\left({\bf x}\right)+\xi({\bf x},t), \quad
 \overline{\xi({\bf x_1},t_1)\xi({\bf x_2},t_2)}=2T
\delta(t_1-t_2)$$
 then the probability $P({\bf x},t)$ to find such a particle at a point
${\bf x}$ of the sample of finite size $L$ will converge to the equilibrium
Gibbs-Boltzmann measure $$P({\bf x},t\to \infty)\to p_{\beta}({\bf
x})=\frac{1}{Z(\beta)}\exp{-\beta V({\bf x})}\,$$ characterized by
the inverse temperature $\beta={1}/{T}$.
The normalization $\int_{|{\bf x}|\le L} p_{\beta}({\bf x}) d {\bf
x}\,=1$ implies the value of the partition function to be given by
\begin{equation}\label{freeendef}
Z(\beta)=\int_{|{\bf x}|\le L} \exp{-\beta V({\bf x})}\, d {\bf
x}\,.
\end{equation}

As obviously $\ln{p_{\beta}({\bf x})}= const -\beta V({\bf x})$
the weights $p_{\beta}({\bf x})$ according to our discussion will be multifractal if the potential $V({\bf x})$ is chosen
logarithmically correlated in space:
\begin{equation}\label{2}
\left\langle V\left({\bf x}_1\right) \, V\left({\bf
x}_2\right)\right\rangle=-\,g^2\ln{\left[\frac{({\bf x}_1-{\bf
x}_2)^2+a^2}{L^2}\right]},\quad a\ll L, \quad {\bf x}\in
\mathbb{R}^N\,,
\end{equation}
where we assumed $|{\bf x}|<L$, and the parameter $a$ stands for a
small-scale cutoff.

According to the general discussion, the multifractal structure of
the Gibbs-Boltzmann measure can be extracted from the knowledge of
moments
\begin{equation}\label{BGmom}
\quad P_q=\int_{|{\bf x}|\le L} p^q_{\beta}({\bf x})\, d {\bf
x}=\frac{Z(\beta q)}{\left[Z(\beta)\right]^q}\sim
L^{-N\tau_q}\quad \mbox{as}\quad L\to \infty\,.
\end{equation}
Identifying $M\sim (L/a)^N$ , the Eqs.(\ref{BGmom}) and
(\ref{freeendef}) imply the following expression for the {\it
typical} exponents $\tau_q$ in terms of the appropriately
normalized free energy of the system
\begin{equation}\label{BGmom2}
\quad \tau_q=|q|\beta{\cal F}(|q|\beta)-q\beta{\cal F}(\beta),
\quad {\cal F}(\beta)=-\lim_{M\to
\infty}\frac{\left\langle\ln{Z(\beta)}\right\rangle}{\beta
\ln{M}}\,.
\end{equation}

As shown in the Appendix A, the most natural random field with logarithmic correlations corresponds to the so-called
Gaussian Free Field (GFF) in two spatial dimensions $N=2$, as well as its one-dimensional subsets.
It is one of the fundamental objects in physics and various issues of its statistics attracted
 a lot of interest recently in conformal field theory, Schramm-Loewner evolution, and two-dimensional quantum gravity,
 see e.g. some discussion in \cite{DuSh}. Technically the problem of extracting the multifractality
exponents $\tau_q$ for the GFF amounts to ability to calculate efficiently
the disorder average of the free energy (\ref{BGmom2}). Such task is in general considered to be one of the most difficult problems
in the statistical mechanics of systems with quenched disorder and we will not be able to perform such calculation explicitly
in $N=2$ GFF case \footnote{Actually, in recent years some sophisticated probabilistic methods
were developed which allowed to address somewhat similar questions for GFF, see e.g. \cite{Daviaud} and the references therein. That development however
goes beyond the remit of the present lectures.}.  Instead, we are going
 to outline such calculation for two particular choices of the models with logarithmically
correlated potentials where such calculation is indeed feasible. The
first model uses a special definition of distances between points in
space and is based on the idea of {\it multiplicative
cascades} which originated in the theory of turbulence, see e.g.
discussion and further references in \cite{multcasc}. In fact, the
model is essentially equivalent to statistical mechanics of directed
polymers on disordered trees studied long ago in the seminal paper
by B. Derrida and H. Spohn \cite{DS}.
 Our second model will use standard Euclidean distances but exploits high
dimensionality of the embedding space: $N\to \infty$. Although the
details of the two  models and the corresponding methods of solution
 may look rather different, there is a general consensus that they address essentially the same physics: the so-called
freezing transition common to all disordered systems with logarithmic correlations. And indeed we shall see that the resulting
multifractality spectrum will be identical. In the final section we will give a short account of recent works on different
aspects of logarithmically correlated potentials.

\section{Statistical mechanics for logarithmically correlated potentials generated by multiplicative cascades}
The construction we are going to describe below can be easily carried out in any spatial dimension, but for simplicity we consider
the one-dimensional case of an interval of length $L$ with the left end at the origin.
With each point $0\le {\bf X}\le L$ of such an interval we can associate an infinite binary string generated by expansion
\begin{equation}\label{cas1}
{\bf X}=L\left(\frac{x_1}{2}+\frac{x_2}{2^2}+\ldots+\frac{x_n}{2^n}+\ldots \right)=(x_1x_2x_3\ldots x_n\ldots)
\end{equation}
where each $x_n$ is either $0$ or $1$. For some numbers the binary string is not unique but
 by choosing the expansion with infinite number of zeroes to the right it can always be made unique (
  e.g. we use for $L/2$ the string $(100 \ldots)$ rather than $(0111\ldots)$). Then for any two points ${\bf X}$ and ${\bf Y}$ in the interval we can introduce the {\it distance function}
defined as $d({\bf X},{\bf Y})=\frac{L}{2^{n+1}}$ where $n$ is the {\it maximal} number of first binary digits shared by ${\bf X}$ and ${\bf Y}$.
For example, if ${\bf X}=(0*****\ldots)$   and ${\bf Y}=(1*****\ldots)$ then $n=0$, hence  $d({\bf X},{\bf Y})=\frac{L}{2}$ (which is obviously the maximal possible distance between the points in the interval), if ${\bf X}=(00****\ldots)$   and ${\bf Y}=(01****\ldots)$ then $n=1$, hence  $d({\bf X},{\bf Y})=\frac{L}{2^2}$, etc.
 One can check that such a function $d({\bf X},{\bf Y})$ indeed satisfies all the axioms for the distances: (i) $d({\bf X},{\bf Y})\ge 0, \forall {\bf X}\ne {\bf Y}$, and $d({\bf X},{\bf Y})=0$ implies ${\bf X}={\bf Y}$ (ii) $d({\bf X},{\bf Y})=d({\bf Y},{\bf X})$ and the triangle inequality (iii) $d({\bf X},{\bf Y})+d({\bf Y},{\bf Z})\ge d({\bf X},{\bf Z})$ for any triple ${\bf X},{\bf Y}, {\bf Z}$.

Now, let us associate with every point ${\bf X}$ an infinite set of random i.i.d. variables
$\phi_k({\bf X}),\,\, k=0,1,2,\ldots,\infty$ with zero mean and variances chosen to satisfy:
\begin{equation}\label{cas2}
\left\langle \phi_k\left({\bf X}\right)\phi_l\left({\bf Y}\right)\right\rangle=2g^2\ln{2}\,\delta_{l,k}\delta_{(x_1x_2x_3\ldots x_k),(y_1y_2y_3\ldots y_k)}
\end{equation}
 where we used the Kronecker symbol: $\delta_{A,B}=1$ for $A=B$ and zero otherwise,  for any two objects $A$ and $B$ of arbitrary nature.
 Finally, with any point ${\bf X}$ of the interval we associate a random potential $V\left({\bf X}\right)$ according to the rule
 \begin{equation}\label{cas3}
V\left({\bf X}\right)=\phi_0\left({\bf X}\right)+\phi_1\left({\bf X}\right)+\ldots=\sum_{k=0}^{\infty}\phi_k\left({\bf X}\right)\,.
\end{equation}
This construction implies for any ${\bf X}\ne {\bf Y}$:
 \begin{equation}\label{cas4}
\left\langle V\left({\bf X}\right)V\left({\bf Y}\right)\right\rangle=\sum_{k=0}^{\infty}\left\langle\phi^2_k\left({\bf X}\right)
\right\rangle=2g^2\ln{2}\,(n+1)\,,
\end{equation}
 where we assumed that the two points ${\bf X}$ and ${\bf Y}$ share precisely $n$ first digits in the binary expansion.
This implies that they are separated by the distance $d({\bf X},{\bf Y})=\frac{L}{2^{n+1}}$, hence the above formula takes the form
  \begin{equation}\label{cas5}
\left\langle V\left({\bf X}\right)V\left({\bf Y}\right)\right\rangle=-2g^2\ln{\frac{d({\bf X},{\bf Y})}{L}}, \quad {\bf X}\ne {\bf Y}.
\end{equation}
We see then that with respect to the chosen distance the constructed random potential is logarithmically correlated in space.
When dealing with logarithmically correlated potentials one has to ensure the proper regularization at small distances, as the logarithm
obviously diverges for  ${\bf X}\to{\bf Y}$. Various regularization schemes are possible, and in the present situation one of the most natural
is to replace continuous space of the interval with a discrete lattice structure.
In the particular case under consideration we introduce a "lattice" of $2^K=M$ sites, each site located at
one of the points ${\bf X}_N=\frac{N}{2^K},\quad N=0,1,2,\ldots, 2^{K}-1$. We can visualise this construction via the {\it tree diagram},
associating the random fields $\phi_{l}({\bf X})$ to every branch of the tree
as sketched in Fig. 6 for $K=3$:
\begin{figure}[h]
\begin{picture}(180,200)(-140,-30)
\put(0,-20){\thicklines\line(1,0){170}} \put(0,150){\thicklines \line(1,0){170}}
\put(0,-20){\thicklines\line(0,1){170}} \put(170,-20){\thicklines \line(0,1){170}}
\put(25,10){\begin{picture}(20,20)
\multiput(60,90)(0,0.75){30}{\circle*{1.5}}
\multiput(0,0)(0.5,0.75){120}{\circle*{1.5}}
\multiput(65,0)(0.5,0.75){33}{\circle*{1.5}}
\multiput(83,26)(0.5,0.75){20}{\circle*{1.5}}
\multiput(35,0)(0.5,0.75){20}{\circle*{1.5}}
\multiput(100,0)(0.5,0.75){20}{\circle*{1.5}}
\multiput(120,0)(-0.5,0.75){120}{\circle*{1.5}}
\multiput(55,0)(-0.5,0.75){55}{\circle*{1.5}}
\multiput(20,0)(-0.5,0.75){20}{\circle*{1.5}}
\multiput(85,0)(-0.5,0.75){20}{\circle*{1.5}}
\put(60,95){ $\phi_0$}
\put(18,65){{\small $\phi_1(0)$}}
\put(80,65){{\small $\phi_1(1)$}}
\put(-10,25){{\tiny $\phi_2(00)$}}
\put(40,25){{\tiny $\phi_2(01)$}}
\put(62,25){{\tiny $\phi_2(10)$}}
\put(110,25){{\tiny $\phi_2(11)$}}
\put(-22,5){{\tiny $\phi_3(000)$}}
\put(20,5){ $...$}
\put(53,5){ $...$}
\put(86,5){ $...$}
\put(120,5){{\tiny $\phi_3(111)$}}
\put(0,-7){{\tiny $0$}}
\put(16,-7){{\tiny $\frac{1}{8}$}}
\put(34,-7){{\tiny $\frac{2}{8}$}}
\put(51,-7){{\tiny $\frac{3}{8}$}}
\put(64,-7){{\tiny $\frac{4}{8}$}}
\put(80,-7){{\tiny $\frac{5}{8}$}}
\put(100,-7){{\tiny $\frac{6}{8}$}}
\put(114,-7){{\tiny $\frac{7}{8}$}}
\end{picture}}
\put(70,130){${\sf{K=3}}$}
\end{picture}
\caption{Lattice of $8$ sites and the corresponding tree diagram associating random fields to every branch of the tree.}
\end{figure}
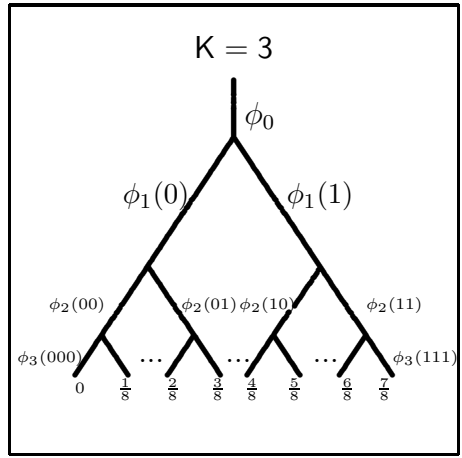

Now we can define the distances in the same fashion as before, but since the maximal number of common digits can be at most $K$ we get
for the variance of the random potential a finite value (cf. (\ref{2})):
  \begin{equation}\label{cas6}
\left\langle V^2\left({\bf X}\right)\right\rangle=2g^2\ln{2}(K+1)\equiv -2g^2\ln{\frac{a}{L}} \,,
\end{equation}
 where we have introduced the effective lattice cutoff given by $a=L/2^{K+1}$. For this regularized lattice version we can now introduce the well-defined Boltzmann-Gibbs weights
 \begin{equation}\label{cas7}
  p_{\beta}({\bf
X}_N)=\frac{1}{Z_K(\beta)}\exp{-\beta V({\bf X}_N)},\quad Z_K(\beta)=\sum_{N=0}^{2^{K}-1}\,\exp{-\beta V({\bf X}_N)}
 \end{equation}
 and  try to calculate the associated free energy $\langle\ln Z_K(\beta)\rangle $, hence to extract the multifractality exponents $\tau_q$, see (\ref{BGmom2}). The value of the potential $V({\bf X}_N)$ associated
with each lattice site ${\bf X}_N$ is obviously obtained by adding all the random fields $\phi({\bf X})$ along the unique path connecting the site to the top level of the tree diagram. This implies the essentially {\it multiplicative} nature of the cascade model for the weight factors $exp\{-\beta V({\bf X}_N)\}$.
The most efficient way to organize calculations amounts to exploiting
such a multiplicative structure combined with the hierarchical organization of the model which is obvious from the tree
diagram decomposition as shown in Fig.7 below. The described structure implies that
\begin{equation}\label{cas8}
 Z_K(\beta)=e^{-\beta \phi_0}\left[ Z_{K-1}^{(L)}(\beta)+ Z_{K-1}^{(R)}(\beta)\right]
 \end{equation}
where $Z_{K-1}^{(L/R)}(\beta)$ corresponds to the left/right-hand subtree of the tree in Fig.7 which is of the depth $K-1$ as reflected in the lower index.

\begin{figure}[h!]
\begin{picture}(160,160)(-140,-20)
\put(0,-20){\thicklines\line(1,0){170}} \put(0,130){\thicklines \line(1,0){170}}
\put(0,-20){\thicklines\line(0,1){150}} \put(170,-20){\thicklines \line(0,1){150}}
\put(25,10){\begin{picture}(20,20)
\multiput(60,90)(0,0.75){30}{\circle*{1.5}}
\multiput(0,0)(0.5,0.75){110}{\circle*{1.5}}
\multiput(65,0)(0.5,0.75){55}{\circle*{1.5}}
\multiput(35,0)(0.5,0.75){20}{\circle*{1.5}}
\multiput(100,0)(0.5,0.75){20}{\circle*{1.5}}
\multiput(120,0)(-0.5,0.75){110}{\circle*{1.5}}
\multiput(55,0)(-0.5,0.75){55}{\circle*{1.5}}
\multiput(20,0)(-0.5,0.75){20}{\circle*{1.5}}
\multiput(85,0)(-0.5,0.75){20}{\circle*{1.5}}
\put(63,98){ $\phi_0$}
\put(-5,60){$Z_{K-1}^{(L)}(\beta)$}
\put(90,60){$Z_{K-1}^{(R)}(\beta)$}

\put(20,5){ $...$}
\put(53,5){ $...$}
\put(86,5){ $...$}
\put(0,-10){$\underbrace{\quad\,\,  \mbox{Left} \quad  }$}
\put(65,-10){$\underbrace{\quad\,\,  \mbox{Right} \quad  }$}
\end{picture}}

\end{picture}
\caption{The tree diagram decomposition leading to recursive relations for the partition function.}
\end{figure}
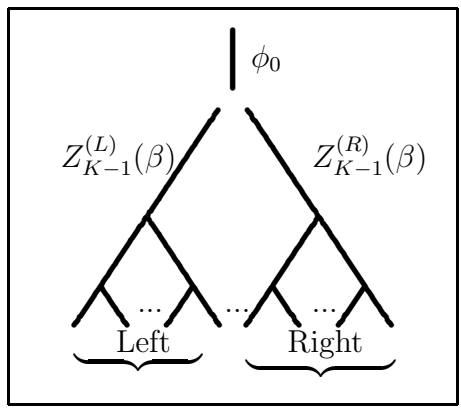

 Note that the fields $\phi({\bf X})$ entering $Z_{K-1}^{(L)}(\beta)$ are statistically independent of those entering $Z_{K-1}^{(R)}(\beta)$.
To make the direct use of the structure of the equation (\ref{cas8}) it is expedient to introduce the generating function
 \begin{equation}\label{cas9}
G_K(p)=\left\langle e^{-p Z_K(\beta)}\right\rangle, \quad p\ge 0\,,
 \end{equation}
which is simply the Laplace transform of the probability density of the partition function. Denoting the probability density
of the distribution for the variable $\phi_0$ with ${\cal P}(\phi_0)$ and exploiting that the variables $Z_{K-1}^{(L)}(\beta)$
and $Z_{K-1}^{(R)}(\beta)$ are independent of each other and identically distributed (i.i.d.) we
arrive at the relation:
\begin{equation}\label{cas10}
\fl G_K(p)=\int {\cal P}(\phi_0)\left\langle e^{-p e^{-\beta \phi_0}[Z^{(L)}_{K-1}(\beta)+Z^{(R)}_{K-1}(\beta)]}\right\rangle\,d\phi_0\equiv
\int {\cal P}(\phi)G^2_{K-1}\left(pe^{-\beta \phi}\right)\,d\phi\,.
 \end{equation}
 Precisely in the same way we can relate $G_{K-1}$ to $G_{K-2}$, etc. in a kind of recursive procedure
 which starts with the obvious initial condition $G_0(p)=e^{-p}$.
 Finally, it turns out that the subsequent analysis becomes more transparent if one
 introduces a new variable $x=-\frac{1}{\beta}\ln{p}\in(-\infty,\infty)$. We arrive therefore at the recursion relations
 \begin{equation}\label{cas11}
  G_l(x)=\int {\cal P}(\phi)G^2_{l-1}\left(x+\phi\right)\,d\phi, \quad l=1,2,\ldots K\,\,\mbox{and}\,\, G_0(x)=e^{-e^{-\beta x}}\,,
 \end{equation}
where we have replaced $ G_l\left(p=e^{-\beta\,x}\right)\to G_l(x)$, with some abuse of notations.

{\bf Note:} If from the very beginning we had considered a tree with arbitrary constant branching $s>1$
instead of the binary tree with $s=2$ the above recursion
would be simply  replaced by
\begin{equation}\label{cas11a}
 G_l(x)=\int {\cal P}(\phi) G^s_{l-1}\left(x+\phi\right)\,d\phi, \quad G_0(p)=e^{-e^{-\beta x}}\,.
 \end{equation}
 where $\left\langle V^2\left({\bf X}\right)\right\rangle=  2g^2(K+1)\ln{s}$ is the variance of the underlying logarithmically correlated potential, cf. ({\ref{cas6}}), and $M=s^K$ is the total number of points in the lattice.

 To understand better the nature of the solution of the above equations in the thermodynamic limit $K\to \infty$ it is instructive to consider the following limiting case for the branching parameter: $s=1+\delta,\,\,\delta\ll 1$ . This implies scaling the variable $\phi$ in such a way that $ <\phi^2>\equiv 2g^2\ln{s}\approx 2g^2\delta$.

 To be specific, one may just wish to use the Gaussian distribution ${\cal P}(\phi)=
 \frac{1}{\sqrt{2\pi\delta}g}\exp{-\frac{\phi^2}{4g^2\delta}}$. Then the right-hand side of (\ref{cas11a}) takes the form
 \[
  \int\frac{1}{\sqrt{2\pi\delta}g}e^{-\frac{\phi^2}{4g^2\delta}}\, G^{1+\delta}_{l-1}\left(x+\phi\right)\,d\phi\equiv
 \int_{-\infty}^{\infty}e^{-\frac{y^2}{2}}\, G^{1+\delta}_{l-1}\left(x+g y\sqrt{2\delta}\right)\,\frac{dy}{\sqrt{2\pi}}
 \]
which after straightforwardly expanding in powers of $\delta$ reduces (\ref{cas11a}) to
\begin{equation}\label{cas11b}
 G_l(x)=G_{l-1}\left(x\right)+\delta\left[G_{l-1}\left(x\right)\ln{G_{l-1}\left(x\right)}+g^2\frac{d^2}{dx^2}G_{l-1}\left(x\right)\right]+
 O(\delta^2)\,.
 \end{equation}
 Thus in such an approximation the function  $G_l(x)$ experiences only small change in one step of iteration: $G_l(x)-G_{l-1}(x)\propto \delta$.
 Introducing to this end the variable $t=l\delta$ and consider it to be continuous in the interval $t\in[0,t_{max}=K\delta\approx \ln{M}]$
 we can replace $G_l(x)\to G(x,t)$ and approximately write to the leading order
$G_l(x)-G_{l-1}(x)\approx \delta\frac{\partial}{\partial t}G(x,t)$. In this approximation the relation (\ref{cas11b}) is replaced by a partial differential equation on the function $G(x,t)$:
\begin{equation}\label{cas12}
\frac{\partial G}{\partial t}=g^2\frac{\partial^2 G}{\partial x^2}+G\ln{G}, \quad G(x,0)=e^{-e^{-\beta x}}\,.
 \end{equation}
 We also note that (i) by its very definition the function $G(x,t)$ satisfies the following conditions:
\begin{equation}\label{cas13}
 0\le G(x,t)\le 1, \quad G(x\to -\infty,t)=0, \quad G(x\to \infty,t)=1\,
 \end{equation}
and (ii) the values $G(x,t)=0$ and $G(x,t)=1$ solves the equation (\ref{cas12}). All these observations are typical for the partial differential
equations having the so-called {\it travelling waves solutions} of the form
\begin{equation}\label{cas14}
\fl G(x,t)=W[x-m(t)], \quad \frac{d}{dt}m(t)\equiv c(t)\to c\,t\quad \mbox{when}\quad t\to \infty, \,\, c=const>0
 \end{equation}
 where the constant $c$ plays the role of the asymptotic velocity of the front propagation, see Fig.8.

\begin{figure}[h!]
 \begin{picture}(120,140)(-120,-20)
\put(0,0){\thicklines\line(1,0){170}} \put(0,100){\thicklines \line(1,0){170}}
\put(0,0){\thicklines\line(0,1){100}} \put(170,0){\thicklines \line(0,1){100}}

\put(25,10){\begin{picture}(20,20)
\put(20,10){\vector(0,1){70}}
\put(-10,20){\vector(1,0){140}}

\multiput(20,60)(15,0){8}{\line(1,0){3}}
\put(8,60){$1$}
\put(10,10){$0$}
\put(24,70){$G(x,t)$}
\put(135,18){$x$}

\qbezier[1000](-20,22)(75,22)(80,45)
\qbezier[1000](80,45)(85,58)(130,58)

\put(85,40){$\rightarrow ct$}

\end{picture}}
\end{picture}
\caption{Sketch of a typical front of the travelling wave solution.}
\end{figure}
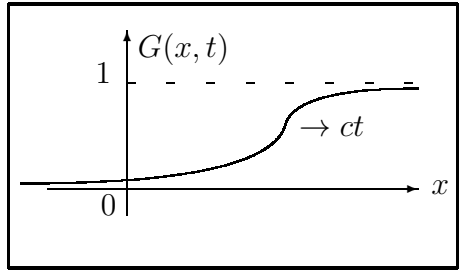

Substituting such a form to (\ref{cas13}), denoting $\tau=x-m(t)$
 (so that e.g. $\frac{\partial G}{\partial t}=-\frac{dW}{d\tau}\frac{d}{dt}m(t)$) we see that the partial differential
  equation in the limit $t\to \infty$ implies an ordinary differential equation for $W(\tau)$ which can be written as:
 \begin{equation}\label{cas15}
g^2\ddot{W}+c\dot{W}+\frac{d}{dW}U(W)=0, \quad \mbox{where}\quad U(W)=\frac{W^2}{2}\left(\ln{W}-\frac{1}{2}\right)
 \end{equation}
where we have introduced the notations $\dot{W}\equiv \frac{dW}{d\tau}$ and $\ddot{W}\equiv\frac{d^2W}{d\tau^2}$.
Obviously, we can interpret the latter equation as the Newtonian equation describing the motion
of a classical particle of mass $g^2$ on the interval of the fictitious "coordinate" $W\in[0,1]$ in fictitious "time" $\tau$
subject to the dissipative force ("friction") $c\dot{W}$ plus the potential force generated by the potential $U(W)$ sketched
 in Fig. 9:

\begin{figure}[h!]
\begin{picture}(120,140)(-120,-20)
\put(0,0){\thicklines\line(1,0){170}} \put(0,100){\thicklines \line(1,0){170}}
\put(0,0){\thicklines\line(0,1){100}} \put(170,0){\thicklines \line(0,1){100}}

\put(25,10){\begin{picture}(20,20)
\put(0,10){\vector(0,1){70}}
\put(0,40){\vector(1,0){140}}

\multiput(75,10)(0,5){6}{\line(0,1){2}}
\put(75,40){\circle*{2}}
\put(0,40){\circle*{2}}
\put(75,10){\circle*{2}}
\put(75,42){$1$}
\put(-10,42){$0$}
\put(5,70){$U(W)$}
\put(130,28){$W$}

\qbezier[1000](0,40)(1,10)(75,10)

\put(35,00){$\mbox{\tiny \bf stable equilibrium}$}
\put(5,55){$\mbox{ \tiny \bf unstable}$}
\put(5,45){$\mbox{\tiny \bf equilibrium}$}
\end{picture}}
\end{picture}
\caption{Sketch of the potential driving the motion of a fictitious overdamped Newtonian particle, see the text.}
\end{figure}
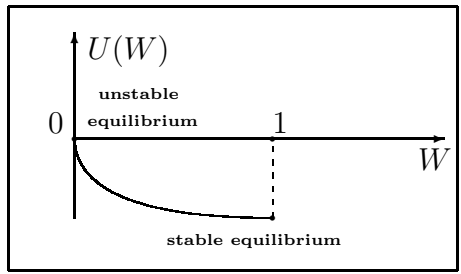

By inspection we see that the position $W=0$ corresponds to the maximum of the potential, hence it is {\sf unstable equilibrium},
and $W=1$ is the {\sf stable equilibrium} (minimum of the potential). As by its physical meaning $W\le 1$ the motion
of such particle must be {\it overdamped},
that is it should approach the stable equilibrium $W=1$ in a monotonic way as $\tau\to \infty$ (i.e. the damping should be strong
enough to avoid oscillations around the stable equilibrium which would bring $W$ out of the physical interval.) To this end, let us
consider in more detail the motion in the vicinity of the stable equilibrium  by expanding: $W=1-v, \,\,v(\tau)\ll 1$ so that to the linear order
$W\ln{W}\approx -v$ and (\ref{cas15}) is reduced to the linear second-order differential equation $g^2\ddot{v}+c\dot{v}+v=0$ whose general solution is given by
\begin{equation}\label{cas16}
v(\tau)=Ae^{\lambda_+\tau}+Be^{\lambda_-\tau},\quad \lambda_{\pm}=-\frac{1}{2g^2}(c\pm\sqrt{c^2-4g^2})
 \end{equation}
To have a non-oscillatory ("overdamped") asymptotic behaviour for $\tau\to \infty$ is only possible for $c\ge 2g$, so that
\begin{equation}\label{cas16a}
v(\tau\to \infty)\approx \left\{\begin{array}{cc} Be^{-\frac{\tau}{2g^2}(c-\sqrt{c^2-4g^2})}, & c>2g
\\ B\tau e^{-\frac{\tau}{g}}, & c=2g\end{array}\right.
 \end{equation}
 To determine the value of $c$ it is natural to recall that according to the definition (\ref{cas9}) $G_K(x)=\left\langle \exp\{-e^{-\beta x} Z_K(\beta)\}\right\rangle$ so that naively expanding for $x\to \infty$ gives
 $G_K(x\to \infty)\approx 1-e^{-\beta x}\left\langle Z_K(\beta)\right\rangle+\ldots $. Using the Gaussian distribution of the
  random potential chosen for the present model one finds
 \begin{equation}\label{cas17}
 \fl \left\langle Z_K(\beta)\right\rangle=M\left\langle e^{-\beta V({\bf X})}\right\rangle=M\exp\left\{\frac{\beta^2}{2}<V^2({\bf X}>\right\}\approx
 Me^{\beta^2g^2\ln{M}}=M^{1+\beta^2g^2}\,,
 \end{equation}
 which implies
 \begin{equation}\label{cas18}
 \fl G_K(x\to \infty)\approx 1-e^{-\beta\tau_{max}}, \,\, \mbox{with}\,\, \tau_{max}=x-ct_{\max},\,\,  t_{\max}\equiv \ln{M} \,\mbox{and}\, c\equiv c(\beta)=\frac{1}{\beta}+\beta g^2\,.
 \end{equation}
 The above formula for the velocity $c=c(\beta)$ ensures the consistency between the asymptotic behaviour
 in (\ref{cas18}) and in (\ref{cas16a}) as for such a choice holds the relation $\beta\equiv\frac{1}{2g^2}(c-\sqrt{c^2-4g^2})$. Moreover, the choice
 is also compatible with the condition for overdamped motion as  $c-2g=(\beta g-1)^2\ge 0$. However there is some subtlety in that formula
  which is most apparent if we follow the function $c(\beta)$ starting from the high-temperature regime $\beta\ll g^{-1}$.
  We see that the wavefront velocity $c(\beta)$ decreases with increasing $\beta$ (decreasing temperature) down to the minimal value $c(\beta=g^{-1})=2g$, and then
  for $T<T_c=g$ starts increasing again, as schematically shown below.

 \begin{picture}(140,140)(-120,-10)
\put(0,0){\thicklines\line(1,0){120}} \put(0,100){\thicklines \line(1,0){120}}
\put(0,0){\thicklines\line(0,1){100}} \put(120,0){\thicklines \line(0,1){100}}

\put(25,10){\begin{picture}(20,20)
\put(0,0){\vector(0,1){80}}
\put(0,10){\vector(1,0){80}}

\put(39,10){\circle*{2}}
\put(0,25){\circle*{2}}
\put(38,0){$g^{-1}$}
\put(-10,5){$0$}
\put(-14,24){$2g$}
\put(-20,75){$c(\beta)$}
\put(80,00){$\beta$}

\qbezier[1000](5,80)(10,30)(40,25)
\qbezier[20](40,25)(60,25)(80,60)

\end{picture}}
\end{picture}

\noindent {\tiny \sf Temperature dependence of the front velocity. The dotted branch is unphysical and should be replaced with the constant value $c=2g$.}

\vskip 0.5cm

A rigorous mathematical analysis of the travelling wave equations by Bramson\cite{bramson}
revealed that such conclusion is however not quite correct. Namely, Bramson proved that for the initial conditions of the type (\ref{cas12})  the actual velocity of the travelling wave front is indeed given by $c(\beta)=\frac{1}{\beta}+\beta g^2$ for $\beta<\beta_c=g^{-1}$, but sticks to the minimal value
 $c_{min}=2g$ {\it everywhere in the low-temperature regime $\beta>\beta_c$}. Such a picture implies, in particular the asymptotic form
 $W(\tau\to \infty)\approx 1-e^{-\beta_c\tau}$, or equivalently the asymptotics
 \begin{equation}\label{cas18a}
  G(x,t)\approx 1-e^{-\beta_c(x-c_{min}t)}, \quad T<T_c=g
 \end{equation}
so that the profile of the function $ G(x,t)$ turns out to be temperature independent ("frozen")
everywhere in the lower-temperature phase. Such behaviour certainly signals of a kind of strong non-analyticity, as e.g. it invalidates
the expansion of the exponent in $\left\langle \exp\{-e^{-\beta x} Z_K(\beta)\}\right\rangle$ which underlay our "naive" analysis.
It is therefore appropriate to call such a drastic change of the behaviour a phase transition, which is known in the literature as the
{\bf freezing transition}.

Qualitatively, the same picture holds generically for an arbitrary branching  $s>1$, that is for the solution of recursive equation (\ref{cas11a}).
Namely, an extension of the above analysis \cite{DS} shows that in the thermodynamic limit $K\to \infty$ the solution takes the form
\begin{equation}\label{cas19}
\fl G_K(x)=W[x-c_\beta K], \quad  c_{\beta}=\left\{\begin{array}{cc}\frac{1}{\beta}\log{\left[s\int {\cal P}(\phi)\,e^{-\beta\phi}\,d\phi\right]},& \beta<\beta_c\\  \frac{1}{\beta_c}\log{\left[s\int {\cal P}(\phi)\,e^{-\beta_c\phi}\,d\phi\right]},& \beta>\beta_c \end{array}\right.
 \end{equation}
where $\beta_c$ is the point at which the function $c_{\beta}$ from the upper line in (\ref{cas19}) has its minimum: $\frac{d}{d\beta}c_{\beta}|_{\beta=\beta_c}=0$. Such a knowledge allows one to calculate our main object of interest,
the mean free energy $-\beta \overline{F}(\beta)=\lim_{K\to\infty}\frac{1}{K}\left\langle\ln{Z_K(\beta)}\right\rangle$. To this end it is convenient to use
the following integral representation for the logarithm:
\begin{equation}\label{logrep}
\ln{Z}=\int_0^{\infty}\left[e^{-p}-e^{-pZ}\right]\frac{dp}{p}\,.
\end{equation}
Remembering the definition of the generating function: $G_K(p)=\left\langle e^{-pZ_K(\beta)}\right\rangle$ and
$G_0(p)=e^{-p}$ and also the relation $p=e^{\beta x}$ we after averaging of (\ref{logrep}) arrive at the important identity:
 \begin{equation}\label{cas20}
 \left\langle\ln{Z_K(\beta)}\right\rangle=\beta\int_{-\infty}^{\infty}\left[G_0(x)-G_K(x)\right]\,dx\,.
 \end{equation}
Inspecting the travelling wave form of the solution (\ref{cas19})  we observe that in the limit $K\gg \frac{1}{\beta c_{\beta}}$
the difference $G_0(x)-G_K(x)$ (sketched in Fig.10) is approximately equal to unity inside the interval $x\in[\frac{1}{\beta}, K\,c_{\beta}]$, and is negligibly small outside:

\begin{figure}
\begin{picture}(120,140)(-120,-20)
\put(0,0){\thicklines\line(1,0){170}} \put(0,100){\thicklines \line(1,0){170}}
\put(0,0){\thicklines\line(0,1){100}} \put(170,0){\thicklines \line(0,1){100}}

\put(25,10){\begin{picture}(20,20)
\put(20,10){\vector(0,1){70}}
\put(-10,20){\vector(1,0){140}}

\multiput(20,60)(15,0){8}{\line(1,0){3}}
\put(8,60){$1$}
\put(10,10){$0$}
\put(24,70){$G_0(x)-G_K(x)$}
\put(135,18){$x$}

\put(28,10){$\beta^{-1}$}
\put(28,20){\circle*{2}}
\put(110,20){\circle*{2}}
\put(110,10){$C_\beta K$}

\qbezier[1000](-20,22)(25,22)(30,45)
\qbezier[1000](30,45)(32,58)(90,58)
\qbezier[1000](90,58)(110,58)(115,35)
\qbezier[1000](115,35)(120,25)(128,25)

\end{picture}}
\end{picture}
\caption{}
\end{figure}

 This immediately produces the simple result for the limiting free energy:
 \begin{equation}\label{cas21}
 -\beta \overline{F}(\beta)=\lim_{K\to \infty}\frac{1}{K}\left\langle\ln{Z_K(\beta)}\right\rangle=\beta\, c_{\beta}
 \end{equation}
with $c_{\beta}$ given by (\ref{cas19}). Remembering $M\approx s^K$ and using the relation (\ref{BGmom2}) for the typical multifractality exponents, we find
\begin{equation}\label{cas22}
 \tau_{q>0}=\frac{1}{\ln{s}}\,\,\beta\,q\,[ c_{\beta}-c_{\beta q}]\,.
 \end{equation}
In particular, for the earlier considered case of the Gaussian distribution ${\cal P}(\phi)=
 \frac{1}{\sqrt{2\pi\delta}g}\exp{-\frac{\phi^2}{4g^2\delta}}$ we find
 \begin{equation}\label{casgau}
  -\beta \overline{F}(\beta)={\ln{s}}\left\{\begin{array}{cc} 1+\frac{\beta^2}{\beta_c^2}, \quad & \beta<\beta_c=1/g
 \\ 2\frac{\beta}{\beta_c},\quad & \beta>\beta_c=1/g\end{array}\right..
 \end{equation}
 We see that the only control parameter for the model is
$\gamma=\beta^2/\beta_c^2$. After a simple calculation using (\ref{cas22}) we recover the multifractality exponents for this case, which we
are going to present only in the range $q>1$:
\begin{equation}\label{typ}
\tau_{q>1}=\left\{\begin{array}{c}(q-1)(1-\gamma q), \quad 0\le \gamma<\frac{1}{q^2}\\
q(1-\sqrt{\gamma})^2, \quad \frac{1}{q^2}<\gamma<1\,\\
0,\,\,\, \quad \gamma>1\,
\end{array}\right.\,.
\end{equation}
 The phenomenon of vanishing of the exponents  $\tau_{q>1}$ in the low-temperature phase $\gamma=\beta^2g^2>1$
  is one of the manifestations of {\it freezing}. It is qualitatively interpreted in terms of
the Boltzmann measure being essentially localised on a few sites
for low enough temperature or strong enough disorder.  The typical
multifractality spectrum corresponding to the above exponents is obtained according to the Legendre transform (\ref{1a})
which gives
\begin{equation}\label{typspectrum}
f(\alpha)=\left\{\begin{array}{c}1-\frac{1}{4\gamma}\left[\alpha-(1+\gamma)\right]^2\quad
 \mbox{for}\quad  \gamma<1\\ -\frac{1}{4\gamma}\left[\alpha^2-4\sqrt{\gamma}\alpha\right]\quad
 \mbox{for} \quad \gamma>1
\end{array}\right.\,,
\end{equation}
where the expression in the first line formally assumes the range of
exponents $\alpha_{-}=(1-\sqrt{\gamma})^2\le \alpha\le
 1+\gamma=\alpha_0$, whereas in second line $0\le \alpha\le 2\sqrt{\gamma}=\alpha_0$.
 The upper bound $\alpha_0$ here corresponds to the point of maximum of $f(\alpha)$ and is related
 to the formal restriction $q>1$ in (\ref{typ}). In fact however it is not difficult to find $\tau_q$ for any $q$ and show that the expressions (\ref{typspectrum}) are valid in a wider range $\alpha\in [\alpha_{-},\alpha_+]$ where the boundary $\alpha_{+}$ is the largest root of  $f(\alpha)= 0$.

\begin{picture}(380,240)(0,-40)
\put(0,-20){\thicklines\line(1,0){370}} \put(0,160){\thicklines \line(1,0){370}}
\put(0,-20){\thicklines\line(0,1){180}} \put(370,-20){\thicklines \line(0,1){180}}
\put(25,10){\begin{picture}(20,20)
\multiput(00,110)(15,0){5}{\line(1,0){3}}
\put(0,110){\circle*{2}}
\put(-8,106){$1$}
\put(-8,20){$0$}
\put(5,126){$f(\alpha)$}
\put(10,10){$(1-\sqrt{\gamma})^2$}
\put(80,10){$(1+\sqrt{\gamma})^2$}
\put(135,18){$\alpha$}
\put(55,48){$\gamma<1$}
\put(30,20){\circle*{2}}
\put(100,20){\circle*{2}}
\put(0,0){\vector(0,1){130}}
\put(0,20){\vector(1,0){130}}
\qbezier[1000](30,20)(65,200)(100,20)
\end{picture}}
\put(80,140){${\sf{Fig.\,\, 11a}}$}
\put(30,0){${\mbox{multifractality spectrum}}$}
\put(30,-10){${\mbox{in the high-temperature phase}}$}

\put(25,10){\begin{picture}(20,20)
\multiput(200,110)(15,0){4}{\line(1,0){3}}
\put(200,110){\circle*{2}}
\put(192,106){$1$}
\put(192,20){$0$}
\put(205,126){$f(\alpha)$}
\put(290,10){$4\sqrt{\gamma}$}
\put(335,18){$\alpha$}
\put(235,48){$\gamma>1$}
\put(200,20){\circle*{2}}
\put(300,20){\circle*{2}}
\put(200,0){\vector(0,1){130}}
\put(200,20){\vector(1,0){130}}
\qbezier[1000](200,20)(250,200)(300,20)
\end{picture}}
\put(280,140){${\sf{Fig.\,\, 11b}}$}
\put(230,0){${\mbox{multifractality spectrum}}$}
\put(230,-10){${\mbox{below freezing temperature }}$}

\end{picture}

  Exploiting the relation (\ref{typ1}) for the typical multifractality spectrum one has to specify the limits of integration over $\alpha$ to be precisely
$\alpha_{-}\le \alpha \le \alpha_+$. Substituting there (\ref{typspectrum}) and calculating the integral by the steepest descent
 method reproduces the values (\ref{typ}) of the quenched exponents, that is $\tau^{typ}\equiv\tau_q$.
 Such a calculation confirms that
 the change of behaviour of the exponent $\tau_q^{typ}$ to linear in $q$ for $\gamma>1/q^2$ is
induced by the dominance of the boundary point $\alpha_{-}$ in the integration
over $\alpha$, in agreement with general discussion after (\ref{typ1}).

Thinking in terms of the multifractality spectrum
 it is also easy to see that the freezing phenomenon at $\gamma>1$ is related to $\alpha_{-}=0$, when the
leftmost end of the curve $f(\alpha)$ hits the vertical axis
precisely at zero level: $f(0)=0$, see Fig.11b.

\section{Statistical mechanics for logarithmically correlated potentials in Euclidean spaces of high dimensionality}

As was discussed in the Introduction, we consider the Gibbs partition function of a classical
particle confined to a spherical box of some finite radius $L$.
We denote the corresponding domain as $\{D_L:\,|{\bf x}|\le L \}$.
As before our main goal is to calculate the
ensemble average of the free energy
\begin{equation}\label{freendef}
F=-\frac{1}{\beta}\,\ln{Z_{\beta}},\quad Z=\int_{D_L} \exp{-\beta V({\bf x})}\,
d {\bf x}\,,
\end{equation}
where $\beta=1/T$ stands for the inverse temperature and $d{\bf x}$ is the standard volume element
in $N-$dimensional Euclidean space. The average
of the logarithm of the partition function is one of the central problems
in the whole physics of disordered systems, and is usually performed with the help of the
so-called replica trick, i.e. the formal identity
\begin{equation}\label{replica}
\left\langle\ln{Z_{\beta}}\right\rangle=\lim_{n\to
0}\frac{1}{n}\ln{\left\langle Z_{\beta}^n\right\rangle},\quad
Z_{\beta}^n=\int_{D_L} e^{-\beta\sum_{a=1}^n V({\bf
x_{a}})}\prod_{a=1}^n d {\bf x}_{a}\,.
\end{equation}

 The random Gaussian-distributed potential $V({\bf x})$
is characterized by zero mean and the covariance specified by the
pair correlation function (\ref{2}). Performing the averaging over the Gaussian disorder in
Eq.(\ref{replica}) according to the formula (\ref{Gauprob1}), we in the standard way arrive at the following
expression:
\begin{equation}\label{replica1}
\left\langle
Z_{\beta}^n\right\rangle=e^{\gamma n\ln{\frac{L}{a}}}\int_{D_L} e^{-\gamma \sum_{a<b}\ln{\left[\frac{({\bf x}_1-{\bf
x}_2)^2+a^2}{L^2}\right]}}\prod_{a=1}^n d {\bf x}_{a}\,,
\end{equation}
where we recall the definition of the main control parameter of the problem: $\gamma=\beta^2g^2$.
 To achieve further progress one has to suggest an efficient way of working with the
resulting multidimensional non-Gaussian integral. To this end one may notice that the integrand in
Eq.(\ref{replica1}) in fact possesses a high degree of invariance:
it depends on $N-$component vectors ${\bf x_a}$ only via
$n(n+1)/2$ scalar products $q_{ab}={\bf x_a}{\bf x_b},\,\, a\le
b$, and is therefore invariant with respect to an arbitrary
simultaneous $O(N)$ rotation of all vectors ${\bf x}_a$. Moreover,
our choice of the integration domain respects this invariance.
To this end, introduce $N\times n$ rectangular matrix $X=({\bf x}_1,...,{\bf x}_n)$
such that the $N-$ component vector ${\bf x}_i$ forms $i-$th column of such a matrix.
Then the matrix $Q=X^TX$ is $n\times n$ positive definite, whose entries are precisely
the scalar products $q_{ab}={\bf x_a}{\bf x_b},\,\, a\le
b$. An efficient method of dealing  with integrals possessing such type of invariance is based on
the fundamental identity
\begin{equation}\label{trans}
\fl \int_{|{\bf x}_1|<L}... \int_{|{\bf x}_n|<L} {\cal
I}\left(X^TX\right)\,d{\bf x}_1\ldots d{\bf x}_n
={\cal C}_{N,n} \int_{D_L^{(Q)}}{\cal I}(Q)\, \left[\det{Q}\right]^{\frac{N-n-1}{2}}\, dQ\,,
\end{equation}
where $\quad {\cal C}_{N,n}=
\frac{\pi^{\frac{n}{2}\left(N-\frac{n-1}{2}\right)}}
{\prod_{k=0}^{n-1}\Gamma\left(\frac{N-k}{2}\right)}$ and we assumed $N\ge n+1$.
The integration domain in the right-hand side is
simply $D_L^{(Q)}=\{Q\ge 0,\, q_{aa}\le L^2,\, a=1,\ldots n\}$,
the volume element is $dQ=\prod_{a\le b} dq_{ab}$.
 The above formula seem to appear originally in \cite{Percus} but has not been much in use before it was independently rediscovered
 in the context of theory of random matrices in \cite{Fyo}. In \cite{FS} it was exploited in the present context.
 Since the relation turns out to be quite useful in a few applications
 we present in the Appendix B its derivation taken from \cite{Fyo} with the purpose of making the notes self-contained.

 Applying such a transformation gives in our case:
 \begin{equation}\label{replica1a}
\left\langle
Z_{\beta}^n\right\rangle={\cal C}_{N,n}e^{\gamma n\ln{\frac{L}{a}}}\int_{D_L^{(Q)}} e^{-\gamma \sum_{a<b}\ln{\left[\frac{q_{aa}+q_{bb}-2q_{ab}+a^2}{L^2}\right]}} \left[\det{Q}\right]^{\frac{N-n-1}{2}}\, dQ\,,
\end{equation}
So far all our manipulations were exact for any spatial dimension, provided $N\ge n+1$. For any
finite $N<\infty$  no further simplifications seem possible, any ways to proceed to analysis
of (\ref{replica1a}) are presently unknown and yet to be found.

The situation is better if we agree to consider the dimension $N$ as one more control parameter and let it to be large: $N\gg 1$.
After appropriate rescaling of
the coupling constant $g\to g\sqrt{N}$  (i.e. $\gamma\to N\gamma$) and also rescaling the integration variables $Q\to \frac{a^2}{2}Q$
we can rewrite the exact expression for the averaged replicated
partition function in the following form
\begin{equation}\label{replica2}
\left\langle Z_{\beta}^n\right\rangle={\cal C}_{N,n} \left(\frac{a^2}{2}\right)^{Nn/2}
e^{N\gamma n^2\ln{\frac{L}{a}}}
\int_{D_Q}
\left(\mbox{det}Q\right)^{-(n+1)/2} e^{- N\Phi_n (Q) }\, dQ
\end{equation}
where
\begin{equation}\label{repham1}
 \Phi_n (Q)=-\frac{1}{2}\ln{(\det{Q})}+\gamma\sum_{a<b}
\ln{\left[\frac{1}{2}(q_{aa}+q_{bb})-q_{ab}+1\right]}
\end{equation}
and $N$ is assumed to satisfy the constraint $N>n$. The final
integration domain $D_Q$ is: $D_Q=\{Q\ge
0,\, q_{aa}\le \,R^2=2L^2/a^2,\, a=1,\ldots n\}$. The form of the integrand in Eq.(\ref{replica2})
is precisely one required for the possibility of evaluating the replicated
partition function in the limit $N\to \infty$ by the multidimensional Laplace
(also known as the "steepest descent" or "saddle-point") method. The effective free energy relevant for extracting
the multifractality is then calculated by replica trick as
(see (\ref{BGmom2}) and (\ref{replica}))
\begin{equation}\label{repfreeen}
\beta {\cal F}(\beta)=-\lim_{M\to
\infty}\frac{\left\langle\ln{Z_{\beta}}\right\rangle}{
\ln{M}}=\lim_{L\to \infty}\frac{1}{\ln{L}}\lim_{n\to
0}\frac{1}{n}\Phi_n (Q)
\end{equation}
where we have replaced $\ln{M}\approx N\ln{L}$, and the entries of the matrix $Q$
should be chosen to satisfy the extremal
conditions: $\frac{\partial \Phi_n(Q)}{\partial q_{ab}}=0$ for
$a\le b$.  This yields, in general, the system of $n(n+1)/2$
equations:
\begin{equation}\label{spp1}
-\left[Q^{-1}\right]_{aa}+\gamma\sum_{b(\ne a)}^n
\left[\frac{1}{2}(q_{aa}+q_{bb})-q_{ab}+1\right]^{-1}=0,\quad
a=1,2,\ldots,n
\end{equation}
and
\begin{equation}\label{spp2}
 -\left[Q^{-1}\right]_{ab}-\gamma
\left[\frac{1}{2}(q_{aa}+q_{bb})-q_{ab}+1\right]^{-1}=0,\quad a\ne b
\end{equation}

 One should also ensure that the solutions to these equations
respects the constraint $q_{aa}\le R^2$ for all $a=1,\ldots,n$
imposed by the presence of the boundaries of the integration domain
$D_Q$, and also the fact of $Q$ being positive definite. However, the above equations obviously
imply
\begin{equation}\label{spp2a}
\left[Q^{-1}\right]_{aa}= -\sum_{b(\ne a)}\left[Q^{-1}\right]_{ab}, \quad \forall a=1,2,\ldots,n\,.
\end{equation}
The above condition ensures that the matrix $Q^{-1}$ must have at least one zero eigenvalue
(which corresponds to the uniform eigenvector with all components equal)
which is obviously inconsistent with constraints on $Q$.
We interpret such a failure as manifestation of the fact that the functional $\Phi_n(Q)$ cannot achieve its extremum {\it inside} the
domain $Q>0, q_{aa}\le R^2$. This means that such an extremum should be looked for at the {\it boundary}
of the domain: $q_{aa}=R^2, \, \forall a=1,2,\ldots,n$.
In turn, it means that when searching for such an extremum we only vary $\Phi_n(Q)$ with respect the off-diagonal entries, and therefore
only have to satisfy the equation (\ref{spp2}).

Our procedure of investigating the equations
(\ref{spp1},\ref{spp2}) in the replica limit $n\to 0$ will follow
the standard pattern suggested by developments in spin glass theory\cite{Parisi}.
We first seek for the so-called "replica symmetric" solution, and
then investigate its stability depending on $\gamma$. When the
replica symmetric solution is found inadequate, it should be replaced
by the hierarchical ("Parisi", or "ultrametric") ansatz for the
matrix elements $q_{ab}$, with various levels of replica symmetry
breaking.

\subsection{Analysis of the model within the
Replica Symmetric Ansatz.}

The Replica Symmetric Ansatz amounts to searching for a solution
to (\ref{spp1}),(\ref{spp2}) within subspace of $n\times n$ symmetric positive definite matrices $Q$ such
that $q_{aa}=q_d=R^2$, for any $a=1,\ldots n$, and $q_{a<b}=q_0$,
subject to the constraints $0<q_0\le R^2$ to ensure positive definiteness. Inverting
such a $Q$ yields the matrix $Q^{-1}$ of the same structure, with the
diagonal entries all equal and given by
\begin{equation}\label{invsym1}
p_d=\frac{R^2+q_0(n-2)}{(R^2-q_0)(R^2+q_0(n-1))}
\end{equation}
and all off-diagonal entries given by
\begin{equation}\label{invsym2}
p_0=-\frac{q_0}{(R^2-q_0)(R^2+q_0(n-1))}
\end{equation}
Note, that
\begin{equation}\label{invsym}
p_d-p_0=\frac{1}{R^2-q_0}
\end{equation}
In the replica limit $n\to 0$ the equations (\ref{spp2}) and (\ref{invsym2}) give in this way  the equation for $q_0$:
\begin{equation}\label{sp1sym}
\frac{q_0}{(R^2-q_0)^2}-\frac{\gamma}{\left(R^2-q_0+1\right)}=0
\end{equation}
It is convenient to define the variable $d_0=R^2-q_0$ satisfying $0\le d_0\le R^2$ and reduce (\ref{sp1sym}) to the simple quadratic equation
$(\gamma+1)d_0^2-d_0(R^2-1)-R^2=0$.  Choosing the solution with $d_0>0$ and remembering that we are actually
interested in the large$-L$ limit $ R^2=2L^2/a^2\gg 1$ we find
\begin{equation}\label{sp1sym1}
  d_0=\frac{1}{2(1+\gamma)}[R^2-1+\sqrt{(R^2-1)^2+4R^2(1+\gamma)}]\approx \frac{R^2}{\gamma+1},
\end{equation}
 Now we should calculate the value of the functional $\Phi_n(Q)$ for the replica symmetric solution. It is easy to show that
  $\det{Q}=(R^2-q_0)^{n-1}[R^2+(n-1)q_0]$, so that in the limit $n\to 0$ we easily find from (\ref{repham1})
\begin{equation}\label{sp1sym2}
\lim_{n\to 0}\frac{1}{n}\Phi_n (Q)=-\frac{1}{2}\ln{d_0}-\frac{\gamma}{2}\ln{(d_0+1)}\approx -(1+\gamma)\ln{L}+O(\ln{a})
\end{equation}
where we again considered the limit $L\gg a$. This shows that the effective free energy (\ref{repfreeen}) is given by
\begin{equation}\label{repfreeensym}
\beta {\cal F}(\beta)=\lim_{L\to \infty}\frac{1}{\ln{L}}\lim_{n\to 0}\frac{1}{n}\Phi_n (Q)=1+\gamma\equiv 1+\beta^2g^2\,.
\end{equation}
This coincides precisely with the high-temperature ($T>T_c=g$, i.e. $\gamma<1$) result for the "cascade model" of the previous section, cf.
(\ref{casgau}), which is valid before the freezing mechanism becomes operative.
Our next goal is to understand how the freezing emerges and is maintained for $T<T_c$.

\subsection{Analysis within the Parisi scheme of the replica symmetry breaking.}

The standard way of revealing the breakdown of the replica-symmetric solution  is to perform a stability analysis
following the pattern of the famous de Almeida-Thouless paper \cite{AT} in the theory of spin glasses, i.e.
 magnetic systems with random interactions.
Such analysis can be straightforwardly done for the present type of system, see Appendix D of the present lectures,
and shows that for a given value of $R$
the replica symmetric solution becomes unstable  for the temperatures $T<T_c=g\frac{R^2-1}{R^2+1}$.
Therefore at low temperatures stable solution will have to be one with a broken symmetry in the replica space.
To derive the corresponding expression for the free energy of our model we will follow a particular heuristic scheme of the
replica symmetry breaking proposed originally by Parisi in the theory of spin glasses, see e.g. \cite{Parisi}, or more recently
 \cite{Dedominicis} \footnote{In recent years the use of the scheme was justified by alternative rigorous mathematical procedures.
 For the model under consideration the corresponding equations
were re-derived recently by  a rigorous methods in \cite{AK}  without any recourse to
the powerful but ill-defined replica trick.}. To make the present set of lectures self-contained
we describe in full detail the structure of the matrix $Q$, the ensuing Parisi function $x(q)$ and the main steps of the derivation
 in Appendix C in full detail\footnote{The Appendix is taken {\it verbatim} from \cite{FS}, but we
used this opportunity to correct the important formula (\ref{criso11}) which appeared in \cite{FS} in a distorted form.}.
Here we just sketch those objects schematically for the convenience of the reader:

\begin{picture}(220,300)(-20,-80)
\put(-20,-19){\thicklines\line(1,0){170}} \put(-20,160){\thicklines \line(1,0){170}}
\put(-20,-19){\thicklines\line(0,1){179}} \put(150,-19){\thicklines \line(0,1){179}}
\put(25,10){\begin{picture}(20,20)

\put(123,152){$n$}\put(38,153){$m_1$}\put(-22,153){$m_2$}\put(-43,153){$m_3$}

\put(-45,60){\thicklines\line(1,0){170}} \put(40,-30){\thicklines \line(0,1){180}}
\put(-45,120){\thicklines\line(1,0){56}} \put(-16,90){\thicklines\line(1,0){56}}
\put(-16,90){\thicklines\line(0,1){59}}\put(12,120){\thicklines\line(0,-1){59}}
\put(40,30){\thicklines\line(1,0){57}}
\put(69,1){\thicklines\line(1,0){56}}
\put(69,1){\thicklines\line(0,1){59}}
\put(97,30){\thicklines\line(0,-1){59}}
\multiput(-44,135)(28,-30){3}{\line(1,0){27}}
\multiput(41,45)(28,-30){3}{\line(1,0){27}}
\multiput(-30,120)(28,-30){3}{\line(0,1){29}}
\multiput(55,31)(28,-30){3}{\line(0,1){29}}
\multiput(48,45.5)(14,-15){6}{\line(0,1){14}}
\multiput(-37,135)(14,-15){6}{\line(0,1){14}}
\multiput(41,53)(14,-15){6}{\line(1,0){14}}
\multiput(-44,142)(14,-15){6}{\line(1,0){14}}
\multiput(80,105)(-85,-90){2}{$ q_0$}
\multiput(18,125)(-45,-45){2}{$ q_1$}
\multiput(102,37)(-45,-45){2}{$ q_1$}
\multiput(-26,140)(-14,-14){2}{$ q_2$} \multiput(2,110)(-14,-14){2}{$ q_2$}
\multiput(30,81)(-14,-14){2}{$ q_2$} \multiput(58,51)(-14,-14){2}{$ q_2$}
\multiput(86,21)(-14,-14){2}{$ q_2$} \multiput(115,-10)(-14,-14){2}{$ q_2$}

\end{picture}}
\put(-20,-45){${\sf{Fig.\,\, 12a}}$}
\put(23,-40){${\mbox{Schematic hierarchical structure of} }$}
\put(23,-50){${\mbox{the matrix } \, Q\,\, \mbox{in Parisi parametrisation}}$}

\put(25,10){\begin{picture}(20,20)
\put(200,110){\circle*{2}}
\put(192,106){$n$}
\put(192,15){$0$}
\put(205,126){$x(q)$}
\put(335,18){$q$}
\put(200,20){\circle*{2}}
\put(300,20){\circle*{2}}
\put(200,110){\thicklines{\line(1,0){20}}}
\put(220,90){\thicklines{\line(1,0){15}}}
\put(220,90){\thicklines{\line(0,1){20}}}
\put(235,75){\thicklines{\line(1,0){25}}}
\put(235,75){\thicklines{\line(0,1){15}}}
\put(280,45){\thicklines{\line(1,0){20}}}
\put(300,35){\thicklines{\line(1,0){20}}}
\put(300,35){\thicklines{\line(0,1){10}}}
\put(317,10){$q_d$}\put(320,20){\circle*{2}}
\put(297,10){$q_k$}
\put(275,10){$...$}
\put(218,10){$q_0$}\put(220,20){\circle*{2}}
\put(233,10){$q_1$}\put(235,20){\circle*{2}}
\put(258,10){$q_1$}\put(260,20){\circle*{2}}
\put(185,90){$m_1$}\put(200,90){\circle*{2}}
\put(185,75){$m_1$}\put(200,75){\circle*{2}}
\put(188,62){\circle*{1}} \put(188,59){\circle*{1}} \put(188,56){\circle*{1}}
\put(265,62){\circle*{2}} \put(270,59){\circle*{2}} \put(275,56){\circle*{2}}
\put(185,45){$m_k$}\put(200,45){\circle*{2}}
\put(192,35){$1$}\put(200,35){\circle*{2}}

\put(200,0){\vector(0,1){130}}
\put(200,20){\vector(1,0){130}}
\end{picture}}
\put(210,-5){${\sf{Fig.\,\, 12b}}$}
\put(252,0){${\mbox{Step-wise Parisi function }}$}
\put(252,-10){${\mbox{for finite integer $n$ }}$}

\end{picture}

We are actually interested in the replica limit $n\to 0$.
According to the Parisi prescription explained in detail in the
Appendix C, in such a limit $x(q)$ becomes {\it non-decreasing} function of the variable $q$
and the system can be fully described in terms of such an object.
The function depends non-trivially on its argument in the
interval $q_0\le q \le q_k$, with $q_0\ge 0$ and $q_k\le q_d$.
Outside that interval the function stays constant:
\begin{equation}\label{outside}
x(q<q_0)=0,\quad \mbox {and}\quad  x(q>q_k)=1.
\end{equation}

 In general, the function $x(q)$ also depends on the increasing
sequence
 of $k$ positive parameters $m_i$ satisfying the following
 inequalities
\begin{equation}\label{parisiseq2}
0\le m_1\le m_2\le \ldots\le m_k\le m_{k+1}=1\,.
\end{equation}
If the number of levels of the Parisi hierarchy $K$ tends to infinity we may think of the function
$x(q)$ as continuous  in the
interval $q_0\le q \le q_k$, with possible jumps at the end of the interval: $q=q_0$ and
$q=q_k$.

As is shown in the Appendix C, in the replica limit the following
identity must hold for any differentiable function $g(q)$:
\begin{equation}\label{traces}
\lim_{n\to 0}\frac{1}{n}Tr\left[
g(Q)\right]=g\left(q_d-q_k\right)+
\int_{0}^{q_k}g'\left(\int_{q}^{q_d}x(\tilde{q})\,d\tilde{q}\right)\,dq\,.
\end{equation}

In particular, for the first term entering the replica functional Eq.(\ref{repham1}) application of the rule
Eq.(\ref{traces}) gives
\begin{equation}\label{freeenergy1}
 \lim_{n\to 0}\frac{1}{n}\left[
Tr\ln{(Q)}\right]=\ln{(q_d-q_k)}+\int_{0}^{q_k}\frac{1}{\int_{q}^{q_d}x(\tilde{q})\,d\tilde{q}}\,dq\,.
\end{equation}
The last term in Eq.(\ref{repham1}) is also easily dealt with in
the Parisi scheme (see Appendix C), where it can be written as
\begin{equation}\label{freeenergy2}
-\gamma\lim_{n\to
0}\sum_{l=0}^k(m_{l+1}-m_{l})\ln{(q_d-q_l+1)}=-\gamma\int_0^{q_d}\ln{(q_d-q+1)}x'(q)\,dq,
\end{equation}
by using explicitly the derivative of the generalized function Eq.
(\ref{xstep}). Using integration by parts and taking into account
the properties Eq.(\ref{outside}) we finally arrive at the
required free energy functional for the phase with broken replica
symmetry
\begin{eqnarray}\label{freebroken}\nonumber
&& \lim_{n\to 0}\frac{1}{n}\Phi_n (Q)=-\frac{1}{2}\left[\ln{\left(
q_d-q_k\right)}+\int_{0}^{q_k}\frac{1}{q_d-q_k+\int_{q}^{q_k}x(\tilde{q})\,d\tilde{q}}\,dq\right]
\\
&&-\frac{\gamma}{2}\left(\ln{(q_d-q_k+1)}+\int_{q_0}^{q_k}\frac{1}{q_d-q+1}\,x(q)\,dq\right)\equiv \phi\{x(q)\}
\end{eqnarray}
 The functional $\phi\{x(q)\}$ should be now extremized with respect to the
non-negative non-decreasing continuous function $x(q)$, whereas as we know the
variable $q_d$ must be fixed to its boundary value $q_d=R^2$. To this end we find it convenient to introduce two
 parameters $d_{min}=R^2-q_k,\, d_{max}=R^2-q_0$ satisfying $0\le d_{min}\le d_{max}\le R^2$ and also to use
 $t=R^2-q$ as the new integration variable simultaneously replacing
 (with some abuse of notations)  $x(q=R^2-t)\to x(t)$. Such a renamed function $x(t)$ is now
 {\it non-increasing} in the interval $t\in[d_{min},d_{max}]$, and satisfies
$x(t<d_{min})=1, \, x(t>d_{max})=0$.

As the result, the above functional assumes
a somewhat simpler form:
\begin{equation}\label{freebroken1}
\fl -2\phi\{x(t)\}=\ln{\left(d_{min}\right)}+\int_{d_{min}}^{R^2}\frac{\,d t}{d_{min}+
\int_{d_{min}}^{t} x(\tilde{t})\,d\tilde{t}}+\gamma\ln{(d_{min}+1)}+\gamma\int_{d_{min}}^{d_{max}}\frac{x(t)dt}{t+1}
\end{equation}
Varying the functional Eq.(\ref{freebroken1}) with
respect to such a function $x(t)$ gives after due manipulations with integrals the expression
\begin{equation}\label{varfree}
\fl -2\delta \phi\{x(t)\}=\int_{d_{min}}^{d_{max}}\,S(t)\,\delta x(t)=0, \quad S(t)=\gamma\frac{1}{t+1}-\int_t^{R^2}\frac{d\tilde{t}}{\left[d_{min}+\int_{d_{min}}^{\tilde{t}} x(\tau)\,d\tau
\right]^2}\,,
\end{equation}
Requiring the variation to vanish therefore amounts to the condition $S(t)=0,\,\, \forall t\in[d_{min},d_{max}]$.
As this obviously implies $\frac{d}{dt}S(t)=0$ we can
differentiate Eq.(\ref{varfree}) once, and immediately get
the equation
\begin{equation}\label{var1}
d_{min}+\int_{d_{min}}^{t}x(\tau)\,d\tau=\frac{t+1}{\sqrt{\gamma}},\quad \forall t\in[d_{min},d_{max}]\,\Rightarrow x(t)=\frac{1}{\sqrt{\gamma}}
\end{equation}

What remains to be determined are the values for parameters
$d_{min}$ and $d_{max}$. To
this end, we substitute the value $t=d_{min}$ into the first of relations
Eq.(\ref{var1}) which shows that
\begin{equation}\label{dmin}
d_{min}=\frac{1}{\sqrt{\gamma}-1}\,.
\end{equation}
Next, we use the condition $S(d_{max})=0$, which in view of (\ref{varfree}) and $x(t)=0$ for $t\in[d_{max},R^2]$  gives the relation
\[
\gamma\frac{1}{d_{max}+1}=\int_{d_{max}}^{R^2}\frac{d\tilde{t}}{\left[d_{min}+\int_{d_{min}}^{d_{max}} x(\tau)\,d\tau\right]^2}\,.
\]
Substituting here the expressions (\ref{var1},\ref{dmin}) yields after a simple algebra the $\gamma-$independent result:
\begin{equation}\label{dmax}
d_{max}=\frac{R^2-1}{2},
\end{equation}
completing the solution.
According to the general procedure the solution makes sense as
long as $d_{min}\le d_{max}$, and using $\sqrt{\gamma}=g/T$ it is easy to check that the condition
can be rewritten as $T\le T_{c}=g\frac{R^2-1}{R^2+1}$ which defines the low-temperature phase of the model for finite $R$,
with the same $T_c$ as follows from the stability analysis (Appendix D).
In the thermodynamic limit $\lim_{R\to \infty} T_c=g$, that is $\gamma_c=1$ as expected.

 The value of the functional at the extremum
  can be easily calculated by substituting $x(t)=\gamma^{-1/2}$ for $t\in[d_{min},d_{max}]$ and $x(t)=0$ for
$t\in [d_{max},R^2]$ into (\ref{freebroken1}) and using (\ref{dmin}) and (\ref{dmax}). This gives after some algebra
\begin{eqnarray}\label{equilibriumfreeglass}
\fl -\phi\{x(t)\}=\sqrt{\gamma}\ln{\frac{R^2+1}{2}}+\frac{\sqrt{\gamma}}{2}-\frac{1}{2}(\sqrt{\gamma}-1)^2\ln{(\sqrt{\gamma}-1)}
+\frac{1}{2}(\gamma-2\sqrt{\gamma})\ln{\sqrt{\gamma}}
\end{eqnarray}
which finally implies in the thermodynamic limit $L\to \infty$ for the effective free energy the value
\begin{equation}\label{repfreeena}
\beta {\cal F}(\beta)=\lim_{L\to \infty}\frac{1}{\ln{L}}\lim_{n\to 0}\frac{1}{n}\Phi_n (Q)=2\sqrt{\gamma}=2\beta g\,.
\end{equation}
In particular it shows that the free energy value in the low-temperature phase is {\it frozen} i.e. given by the temperature-independent
constant ${\cal F}(\beta)=2g$. This fully corroborates the picture obtained in the framework of logarithmic cascades of the previous section, see (\ref{casgau}).

Before finishing this section it makes sense to discuss in more detail the picture
associated with the freezing transition which manifests itself via the spontaneous breakdown of
replica symmetry. The general interpretation of the freezing
 below $T_c$ is that the partition function becomes dominated by a finite number of
 sites where the random potential is particularly low, and where the particle ends up spending most of its time
 \cite{MB}. For a more quantitative description of the particle localization, useful in the following,
it is natural to employ the overlap function defined as the mean
probability for two {\it independent} particles placed in the same
random potential
 to end up at a given distance to each other. Denoting the
 scaled Euclidean distance (squared) between the two points in the sample as
 ${\cal D}$, and employing the Boltzmann-Gibbs equilibrium measure $p_{\beta}({\bf
x})=\frac{1}{Z(\beta)}\exp{-\beta V({\bf x})}$ the above
probability in thermodynamic equilibrium should be given by
 \begin{equation}\label{overlap}
 \pi({\cal D})=
 \left\langle\int_{|{\bf x}_1|<L}\,d{\bf x}_1\,p_{\beta}({\bf
x}_1)\int_{|{\bf x}_2|<L}d{\bf x}_2\,p_{\beta}({\bf x}_2)\,
 \delta\left({\cal D}-\frac{1}{2}|{\bf x}_1-{\bf
 x}_2|^2\right)\right\rangle_{V}\,
\end{equation}
where again $\delta$ denotes the Dirac's
$\delta$-function. The disorder averaging in (\ref{overlap}) can
be calculated following the same standard steps of the replica
approach as the free energy itself ( see Appendix A of \cite{FB1}).
 With the function $\pi({\cal D})$ in hand we can ask, in particular
  what is the probability for the particle in logarithmically correlated potential
   to end up at ${\cal D}=O(a^2)$, i.e.  at a distance of order of
  the small cutoff scale. The answer turns out to be zero in the high-temperature phase
   $T > T_c$, confirming the particle delocalization
   over the sample. In contrast, in the low-temperature phase $T <
T_c$ the probability is finite: $\pi\left(O(a^2)\right)=1-T/T_c$, since both particles can be trapped
by one and the same, or nearby favorable, deep minima.
At a formal level such a behaviour is directly related to the shape of the function
$x(t)$ which in our case turned out to be rather simple and consisting of three flat regions (see Fig. 13b):
\begin{equation}\label{RSB1}
\fl x(0<t<d_{min})=1, \quad x(d_{min}<t<d_{max})=\gamma^{-1/2}, \quad x(d_{max}<t<R^2)=0\,.
\end{equation}
This essentially means that from the very beginning we could restrict ourselves to the first non-trivial level $k=1$ of
 the Parisi hierarchical scheme, see Eq.(\ref{parisiseq}, \ref{parisiseq1}) instead of
 assuming the most general Parisi scheme for $Q$ at the outset of our procedure. Such a simplified form (see Fig. 13a) of $Q$  below the transition
 is typical for the random energy models and is known in the literature as 1-step RSB scheme, see e.g. \cite{Dedominicis}.
 The equilibrium values of the parameters $q_0,q_1$ and $m_1\equiv m$ found from directly extremizing the corresponding functional $\frac{1}{n}\Phi_n (Q)|_{n\to 0}$ (or equivalently from solving the equations (\ref{spp2}))
  are given by
  \begin{equation}\label{1step}
  q_0=\frac{R^2+1}{2},\, q_1=R^2-\frac{1}{\sqrt{\gamma}-1}\,,\,\,\mbox{ and}\quad
m=\frac{1}{\sqrt{\gamma}}\,.
\end{equation}
 These values fully agree with those of the function $x(t)$ obtained from the general Parisi Ansatz.
 Finally, in the Appendix D we discuss stability of the 1-step RSB solution for the logarithmic potential, and find it is
 actually {\it marginally stable} everywhere in the low-temperature phase. The latter feature is usually associated with the infinite-step Parisi Ansatz, see e.g. \cite{Dedominicis}. This is another manifestation of the fact that the logarithmic case is very special and shares both features of the full-scale infinite and 1-step replica symmetry breaking.

\begin{picture}(200,300)(-20,-60)
\put(-20,-19){\thicklines\line(1,0){170}} \put(-20,160){\thicklines \line(1,0){170}}
\put(-20,-19){\thicklines\line(0,1){179}} \put(150,-19){\thicklines \line(0,1){179}}
\put(25,10){\begin{picture}(20,20)

\put(123,152){$n$}\put(10,152){$m$}\put(-43,152){$1$}

\multiput(-45,90)(57,-60){2}{\thicklines\line(1,0){56}}
\multiput(13,90)(57,-60){2}{\thicklines\line(1,0){56}}
\multiput(-45,90.5)(57,-60){3}{\thicklines\line(0,1){60}}
\multiput(12,90)(57,-60){2}{\thicklines\line(0,1){59}}
\multiput(-43,140)(18,-18){3}{$R^2$}\multiput(-33,132)(18,-18){3}{\circle*{2}}
\multiput(13,80)(18,-18){3}{$R^2$}\multiput(23,72)(18,-18){3}{\circle*{2}}
\multiput(73,20)(18,-18){3}{$R^2$}\multiput(83,12)(18,-18){3}{\circle*{2}}

\multiput(80,105)(-85,-90){2}{$ q_0$}
\multiput(-5,130)(-30,-30){2}{$ q_1$}
\multiput(50,70)(-30,-30){2}{$ q_1$}
\multiput(110,10)(-30,-30){2}{$ q_1$}

\end{picture}}
\put(-20,-45){${\sf{Fig.\,\, 13a}}$}
\put(22,-40){$\mbox{Structure of the matrix } Q\,\, \mbox{in}$}
\put(22,-50){$\mbox{1-step Replica Symmetry Breaking}$}

\put(25,10){\begin{picture}(20,20)
\put(200,110){\circle*{2}}
\put(192,106){$1$}
\put(192,15){$0$}
\put(205,126){$x(t)$}
\put(335,18){$t$}
\put(200,20){\circle*{2}}
\put(300,20){\circle*{2}}
\put(200,110){\thicklines{\line(1,0){30}}}

\put(230,75){\thicklines{\line(1,0){70}}}
\put(300,20){\thicklines{\line(1,0){20}}}

\put(300,20){{\line(0,1){55}}}
\put(230,75){{\line(0,1){35}}}

\put(320,10){$R^2$}\put(320,20){\circle*{2}}
\put(297,10){$d_{max}$}
\put(228,10){$d_{min}$}\put(230,20){\circle*{2}}
\put(184,73){$\frac{1}{\sqrt{\gamma}}$}\put(200,75){\circle*{2}}

\put(200,0){\vector(0,1){130}}
\put(200,20){\vector(1,0){130}}
\end{picture}}
\put(210,-5){${\sf{Fig.\,\, 13b}}$}
\put(252,0){$\mbox{Parisi function for }$}
\put(252,-10){$\mbox{the logarithmic model}$}
\put(252,-20){$\mbox{ with 1-step RSB }$}
\end{picture}

\section{Summary, Historical background and Recent Extensions}
In this set of lectures we have addressed in some detail the spatial structures of
the Boltzmann-Gibbs measure describing a single particle that thermally equilibrated under a random potential with logarithmic correlations.
We have been able to calculate the multifractality spectrum of the measure and revealed the associated freezing transition
by analysing the ensemble-averaged free energy in two special cases by two complementary methods.
The first model introduced logarithmic correlations via employing the hierarchical "multiplicative cascades" construction
and associated definition of the distance function. This way allowed us to perform the analysis of the freezing transition in the framework
of a certain travelling wave equation satisfied by an appropriately defined generating function of partition function moments.
In the second case the spatial dimension $N$ of the system was assumed to be large
 which helped to employ the replica trick combined with the steepest descent method, and to relate freezing to
the phenomenon of spontaneous replica symmetry breaking. In both cases the resulting free energy appears to be given by
essentially the same expression.

As the present day understanding of freezing and related phenomena has already a history of almost
thirty years, it is certainly useful to be aware of a broader context of the problem under consideration.
To this end it is appropriate to mention that an extreme "toy model" case of the problem in hand is
represented by the famous Random Energy Model (REM) by Derrida
 where the freezing phenomenon was discovered and investigated for the first time \cite{REM,GD89}.
 The REM in some loose sense can be looked at as a limiting
"zero-dimensional" $N=0$ case of the model we studied elsewhere in this set of lectures.
It amounts essentially to replacing the logarithmically correlated random potential by a
collection of $M$ {\it uncorrelated} Gaussian variables with the variances chosen to be scaled
 with $M$ in the same way as in the logarithmic case: $<V_i^2>=2g^2\ln{M}$.
 REM is simple enough to allow explicit calculation of the free energy by direct counting of degrees of freedom,
  and the result essentially coincides with (\ref{casgau}).
 A very informative account of the REM problem can be found in the fifth chapter of \cite{Dedominicis}.

Understanding quantitatively the generic statistical-mechanical behaviour of disordered systems for finite
$N$ is notoriously difficult, and even the simplest cases like our single-particle model still present considerable challenges.
 To this end we first need to mention a general
attempt of investigating such model for finite dimensions $N<\infty$ in the thermodynamic limit $L\to \infty$ undertaken
in an insightful paper by Carpentier and Le Doussal \cite{CLD}.
 The approach of Carpentier and Le Doussal
was based on applying a kind of real-space renormalisation group treatment to the free energy distribution.
The authors concluded that for finite spatial dimensions the model with logarithmically correlated
potetial is really distinguished among others of similar kind. Namely, if
correlations of the random potential {\it grow faster} than logarithm with the distance, then in the thermodynamic
limit the corresponding Boltzmann-Gibbs measure turn out to be always {\it localised}  at any  temperature $T<\infty$.
At the same time, if the correlations {\it decay} to zero for large separations (such potentials are natural to call
"short-ranged") than the  Boltzmann-Gibbs measure turns out to be always {\it trivially extended} at any
positive temperature $T>0$. And only for the marginal situation of logarithmic correlations the true REM-like freezing transition
indeed happens at some {\it finite} $T=T_c>0$, at any dimension $N\ge 1$. Indeed, for that case the renormalisation group yielded
after some clever albeit not fully controlled approximations a kind of travelling wave equation for
 the generating function, akin to (\ref{cas12}).
Fortunately, the logarithmic growth is not at all an academic
oddity. The paper of Carpentier and Le Doussal can be warmly recommended for describing the
present model in a broad physical context and elucidating its
relation to quite a few other interesting and important physical
systems, as e.g. quantum Dirac particle in a random magnetic field \cite{2d},
and  directed polymers on trees with disorder \cite{DS}. The latter works
played the fundamental role in advancing the understanding of the freezing transition.
Our presentation in the Section 2 is actually based on an adaptation of material from \cite{2d} and \cite{DS},
with the pedagogic example of branching tending to unity inspired by \cite{Saakian}.

Another line of research which deserves mentioning was pursued recently in \cite{FB1}
where it was revealed that the picture of potentials with short-ranged,
long-ranged, and logarithmic correlations presented in \cite{CLD} is still
incomplete, and misses a rich class of possible behaviour that
survives in the thermodynamic limit $L \to \infty$. Namely, given
any increasing function $\Phi(y)$ for $0<y<1$, it was suggested to
consider Gaussian random potentials whose two-point correlation functions (covariances) take the following scaling
form
\begin{equation}\label{scalingln}
\left\langle V\left({\bf x}_1\right) \, V\left({\bf
x}_2\right)\right\rangle=-2 \ln{L}\,\,
\Phi\left(\frac{\ln{\left[({\bf x}_1-{\bf
x}_2)^2+a^2\right]}}{2\ln{L}}\right),\quad a\ll L, \quad {\bf x}\in
\mathbb{R}^N\,
\end{equation}
which generalizes our (\ref{2}). Actually, the above expression gives back (\ref{2}) for the special
case $\Phi(y)=g^2(y-1)$. As shown in \cite{FB1} the potential with the covariance (\ref{scalingln})
 can be constructed by a superimposing
several logarithmically correlated potentials of the type
(\ref{2}) with different cutoff scales $a_i$, and allow those
cutoff scales to depend on the system size $L$ in a power-law way:
$a_i\sim L^{\nu_i},\, 0<\nu_i<1$ .

The equilibrium statistical mechanics of such system
in the limit $N\to \infty$ and $L\to \infty$ turns out to be precisely
equivalent to that of the celebrated Derrida's
Generalized Random Energy Model (GREM) see \cite{Bovier} and references therein. Namely, the system experiences
a kind of freezing transition at the critical temperature $T_{c}=\sqrt{\Phi'(1)}$.
Below this temperature the equilibrium free energy turns out to be in the thermodynamic limit $L\to \infty$

\begin{equation}\label{freeenfin}
 -{\cal F}(T)=
T\nu_*(T)+\frac{\left[\Phi(\nu_*)-\Phi(0)\right]}{T}
+2\int_{\nu_*}^1\sqrt{\Phi'(y)}\, dy \,, \quad 0\le T\le
T_{c}\,,
\end{equation}
where the parameter $\nu_*$ is related to the temperature $T$ via
the equation $T^2=\Phi'(\nu_*)$. For $T>T_{c}$ the free energy is instead given by
\begin{eqnarray}\label{freeensyma}
 -{\cal F}(T)=T+\frac{\left[\Phi(1)-\Phi(0)\right]}{T}\,.
\end{eqnarray}
 Using the two-point probability defined in (\ref{overlap}) these expressions for the free energy can be given a clear
interpretation as describing a continuous sequence of "freezing
transitions" which start at $T_c$ and continue at all lower temperatures, with freezing happening on smaller and
smaller spatial scales with decreasing temperature \cite{FB1}. This is related also to the nature of the replica symmetry breaking, which
requires for its description the full infinite sequence $K\to \infty$ of hierarchy levels in the Parisi scheme of Appendix $C$.
Such a rich picture results in a more complicated multifractality spectrum $f(\alpha)$
which in contrast to (\ref{typspectrum}) is in general non-parabolic. However, it is appropriate to mention that
the Boltzmann-Gibbs probability measures generated by the random potentials described in (\ref{scalingln}) are rather peculiar, as for any
non-linear function $\Phi(y)$ they
do not satisfy the standard {\it spatial self-similarity} property (\ref{powerlaw}). Instead, it is easy to check that the exponents
$y(q,s)$ and $z(q,s)$ governing the spatial decay of correlations between weights in (\ref{powerlaw}) will be non-trivial functions of the variable
$\frac{\ln{|{\bf x}_1-{\bf x}_2|}}{\ln{L}}$ rather than simple constants. In this way, the exponents governing the decay of correlations for
two points separated by the distance, say, $|{\bf x}_1-{\bf x}_2|\sim L^{1/2}$ will be different from those separated by, say, $|{\bf x}_1-{\bf x}_2|\sim L^{1/3}$. Though such behaviour is certainly not prohibited by first principles, it remains to be seen whether random multifractal measures with such peculiar spatial structure could appear in interesting applications in physics or other sciences.

Although our lectures were centered around the notion of the multifractality spectrum, there is a different, and in essence deeper
aspect of the freezing transition which attracted considerable research interest recently: the issue of the {\it extreme value statistics}
\cite{BM,CLD,FB2,FLDR}. This goes beyond the calculation of the ensemble-averaged value for the free energy $F=-T\ln{Z}(\beta)$,
but aims to describe precise form of the fluctuations around that mean value. Technically it amounts to our ability to calculate the shape of
 the generating function $G(p)$ defined in (\ref{cas9}) in much finer detail (note that in the context of calculating
 typical multifractality exponents actual form of that function appeared to large extent irrelevant).
 As $lim_{T\to 0}F=\min_{{\bf x}}V({\bf x})$ it is obvious that at low enough temperatures the free energy fluctuations are
 dominated by the distribution of the deepest minimum of the random potential in a given sample.
Classifying possible types of extreme value statistics for strongly correlated random variables is an open problem in probability theory with many important applications in natural sciences and beyond, see \cite{CLD} and the references therein.
In particular, it was argued in \cite{CLD} that logarithmically correlated potentials represent a new universality class for extreme value statistics, and recent works \cite{FB2,FLDR} on extremes of the two-dimensional Gaussian free field
(see definition of this important object in Appendix A1 below) along various curves further substantiated that claim.
Another aspect of the problem which certainly deserves to be mentioned here are
intriguing but so far poorly explored connections to two-dimensional quantum gravity models as noticed in
 \cite{Liouville}, discussed in \cite{CLD}, and most recently in  \cite{FLDR}. Some speculations about relevance
 of the REM-type models in the string theory context can be found in \cite{strings}.

Finally, let us mention that there exists a completely different source of interest in multifractal random processes $\&$ measures with
logarithmic correlations motivated by growing applications in financial mathematics, see e.g. \cite{BMD}, \cite{Ostrovsky} for
the background information and further references. Although the questions addressed there are formally rather different,
one can recognize a common mathematical structure. It is therefore natural to expect a fruitful merger of the two lines of research
in the nearest future.

\vskip 0.5cm

{\bf Acknowledgements.} My understanding of some aspects of the subject of the present lectures was informed
by discussions on various occasions with Jean-Philippe Bouchaud, Pierre Le Doussal and Alexander Mirlin. I am grateful to them
as well as to Hans-Juergen Sommers and Alberto Rosso for collaboration at various stages, and to Ferdinand Evers for kindly providing
picture Fig. 1 for the present notes.

\appendix

\section{Elementary facts about Gaussian integrals and processes, the steepest descent method, and the Gaussian free field}
The fundamental role in applications is played by the standard Gaussian integral
\begin{equation}\label{Gau1}
\int_{-\infty}^{\infty}e^{-\frac{a}{2}y^2+b\,y} \, \frac{dy}{\sqrt{2\pi}}=\frac{1}{\sqrt{a}}\,e^{\frac{b^2}{2a}}, \quad Re{(a)}>0, \, \forall b
\end{equation}
Suppose now we are interested in finding the asymptotic behaviour for large values of a parameter $N$ of the following integral
\begin{equation}\label{sp1}
\int_{y_1}^{y_2}e^{-NF(y)}\phi(x)\,dy, \quad N\gg 1
\end{equation}
where $F(y)$ and $\phi(y)$ are some given infinitely differentiable functions.
It is clear that if the function $F(y)$ is monotonically increasing/decreasing in the interval $y\in[y_1,y_2]$, then
the integral will be dominated by the vicinity of the left/right end of the interval,
and for getting the leading asymptotics it is therefore enough
to expand $F(y)$ around the corresponding point up to the linear term only.
For example for $F'(y)>0, \forall y\in[a,b]$, we write $F(y)\approx F(y_1)+F'(y_1)(y-y_1)+\ldots$ which gives
\begin{equation}\label{sp2}
\int_{y_1}^{y_2}e^{-NF(y)}\phi(y)\,dy \approx \frac{1}{N F'(y_1)}\,e^{-NF(y_1)}\phi(y_1)+O(N^{-2})
\end{equation}
where we assumed that generically $\phi(y_1)\ne 0$ (otherwise one has also to expand $\phi(y)$ around $y_1$,
which will change the result slightly).

Similarly, if the function $F(y)$ has a single {\it maximum} in some point $y_0$ inside the interval,
then subdividing the integration domain into two subintervals $y\in[y_1,y_0]$ and $y\in[y_0,y_1]$
 we can apply the above consideration to each of the new intervals.
For example, if $F(y_1)<F(y_2)$ we have the same asymptotics as above in (\ref{sp2}), whereas for $F(y_1)>F(y_2)$ we have
\begin{equation}\label{sp3}
\int_{y_1}^{y_2}e^{-NF(y)}\phi(y)\,dy \approx \frac{1}{N F'(y_2)}\,e^{-NF(y_2)}\phi(y_2)[1+O(1/N)]
\end{equation}
Finally, the most interesting case arises if $F(y)$ has a single {\it minimum} in some point $y_0\in[y_1,y_2]$, that is
$F'(y_0)=0$ and $F''(y_0)>0$. In such a case the integral will be obviously dominated by the vicinity of the point of minimum,
around which we can therefore expand as $F(y)\approx F(y_1)+\frac{F''(y_1)}{2}(y-y_1)^2+\ldots$.
Substituting this approximation back to the integral and again assuming
that generically $\phi(y_0)\ne 0$ we find after application of (\ref{Gau1}) with $a=F''(y_0)$, the asymptotics
 \begin{equation}\label{sp4}
\int_{y_1}^{y_2}e^{-NF(y)}\phi(y)\,dy \approx \sqrt{\frac{2\pi}{N F''(y_0)}}\,e^{-NF(y_0)}\phi(y_0)[1+O(1/N)]
\end{equation}
These formulae represent the essence of the steepest descent (a.k.a. the Laplace) method
of asymptotic evaluations of integrals.

All the formulae can be naturally extended to the multivariable case. The multivariable generalisation of
the Gaussian integral is given by
\begin{equation}\label{Gaumult}
 \int\ldots \int e^{-\frac{1}{2}\sum_{ij}A_{ij}y_iy_j+\sum_ib_iy_i} \frac{dy_1\ldots dy_n}{(2\pi)^{n/2}}=
\frac{1}{\sqrt{\det{A}}}\,\, e^{\frac{1}{2}\sum_{ij}[A^{-1}]_{ij}b_ib_j}
\end{equation}
where $n\times n$ matrix $A$ is assumed to be real symmetric $A_{ij}=A_{ji}, \, \forall i,j$
and positive definite, i.e. all its eigenvalues $\lambda_i$ are  positive. Then the inverse matrix $A^{-1}$ is well-defined
and the determinant $\det{A}=\prod_{i=1}^n\lambda_i\ne 0$. In fact introducing the scalar product for two vectors as
$({\bf y},{\bf x})=\sum_{i}y_ix_i$ the quadratic form in the exponential
can be written as $\sum_{ij}A_{ij}y_iy_j\equiv ({\bf y},\,A{\bf y})$. The matrix is positive definite iff  $({\bf y},\,A{\bf y})>0,\,\forall {\bf y}$
\footnote{In fact the domain of validity of the formula (\ref{Gaumult}) is broader, and  allows the matrix $A$ to have complex eigenvalues with
positive real parts}.

The analogue of (\ref{sp4}) has the form
\begin{equation}\label{sp4mult}
\fl \int\ldots \int e^{-NF(y_1,,\ldots y_n)}\phi(y_1,\ldots y_n)\,dy_1\ldots dy_n\approx \sqrt{\frac{(2\pi)^n}{N^{n} \det{\delta_2F|_{min}}}}
 \,e^{-NF(y_1,,\ldots y_n)}\phi(y_1,,\ldots y_n)|_{min}
\end{equation}
where we assumed that the function $F(y_1,,\ldots y_n)$ has a single minimum at some point, and
$\delta_2F|_{min}$ is the $n\times n$ Hessian matrix $(\delta_2F)_{ij}=\frac{\partial^2}{\partial y_i \partial y_j}F(y_1,,\ldots y_n)$
evaluated at the point of minimum of $F$.

Let us clarify the probabilistic meaning of the integral (\ref{Gaumult}).
Suppose that $n$ real variables $v_1,\ldots, v_n$ are Gaussian-distributed, that is their normalized
joint probability density of the vector ${\bf v}=(v_1,\ldots, v_n)$ is
given by ${\cal P}(v_1,\ldots, v_n)= e^{-\frac{1}{2}({\bf v},\,A{\bf v})}\sqrt{\frac{\det{A}}{(2\pi)^n}}$ with some positive definite
matrix $A_{ij}$.
Denoting the averaging over such a distribution with the angular brackets $\langle\ldots \rangle$
we can rewrite (\ref{Gaumult}) for any given vector ${\bf b}=(b_1,\ldots, b_n)$ as
\begin{equation}\label{Gauprob}
\left\langle e^{({\bf b},\,{\bf v})} \right\rangle=e^{\frac{1}{2}({\bf b},\,A^{-1}{\bf b})} \quad \Rightarrow \quad  \langle v_i\rangle =0\,\, \mbox{and}\,\,
\left\langle v_iv_j \right\rangle=[A^{-1}]_{ij}, \,\, \forall i,j
\end{equation}
where the identities for the mean value and the pair correlation functions (a.k.a. covariances) immediately follow after expanding in
the Taylor series with respect to $b_i$.

\subsection{Gaussian random fields: "massive" vs "free".}

The last expression is the basis for discussing properties of {\it random processes} (i.e. random functions $V(x)$ of a single real variable $x$)
which are a particular case of {\it random fields} representing random functions $V({\bf x})$ of $N-$dimensional vector ${\bf x}=(x_1,\ldots,x_N)$.
The field  $V({\bf x})$ is called Gaussian if for any choice of the number $n=1,2,\ldots,\infty$ of points ${\bf x}_1, {\bf x}_2,\ldots, {\bf x}_n$  in the space the joint probability density ${\cal P}(v_1,\ldots, v_n)$
of $n$ values of the field in those points, that is $v_1=V({\bf x}_1), v_2=V({\bf x}_2),\ldots, v_n= V({\bf x}_n)$ are given by a Gaussian distribution with some matrix $A_{ij}$. Such random field is uniquely determined by the two-point correlation function (the covariance)
$\langle V({\bf x}_1)\,V({\bf x}_2)\rangle=f({\bf x}_1,{\bf x}_2)$ in terms of which the analogue of (\ref{Gauprob}) reads
\begin{equation}\label{Gauprob1}
\left\langle \exp{\left[{\int b({\bf x})V({\bf x})\,d^N{\bf x}}\right]} \right\rangle=\exp{\left[\frac{1}{2}\int\int f({\bf x}_1,{\bf x}_2) b({\bf x}_1)
\,b({\bf x}_2)\,\, d^N{\bf x}_1 d^N{\bf x}_2\right]}
\end{equation}
for any suitable function $b({\bf x})$. If we define the scalar product of any two functions $a({\bf x})$ and $b({\bf x})$ in the standard way as
$({\bf a},{\bf b})=\int a({\bf x})b({\bf x})\,d^N{\bf x}$, we see that the quadratic form in the exponential of the right-hand
side is $({\bf a},\, \hat{F} {\bf b})$, where the linear integral operator $\hat{F}$ is defined via the kernel $f({\bf x}_1,{\bf x}_2)$.
If one then defines the inverse operator as $\hat{A}=\hat{F}^{-1}$ , the joint probability density of the random field  $V({\bf x})$
can be symbolically written using the scalar product as
 \begin{equation}\label{Gauprob2}
{\cal P}\left[V({\bf x})\right]=\frac{1}{{\cal N}}\,\exp{\left[-\frac{1}{2}(V,\hat{A}V)\right]},
\end{equation}
where ${\cal N}$ is the suitable normalisation constant.

To illustrate the latter approach, we briefly describe the paradigmatic example of the
{\it massive Gaussian field} in $N$ dimensions which is of importance for us here, and is also
 central for the modern theory of phase transitions. The probability of a given configuration $V({\bf x})$ of such field
 is given by (\ref{Gauprob2}) with the quadratic form defined by
 \begin{equation}\label{Gaumass}
\fl (V,\,\hat{A}V)=\int\left(m^2 V^2({\bf x})+
\kappa^2\,[\nabla V({\bf x})]^2\right)\,d^N{\bf x}\equiv \int V({\bf x})\left[m^2 -
\kappa^2\,\Delta \right]\,V({\bf x})\,d^N{\bf x}
\end{equation}
where the "mass" $m$ and the "stiffness" $\kappa$ of the field are two parameters, $\nabla$ is the gradient operator and
$\Delta$ is the Laplacian: $\Delta=\sum_{i=1}^n\frac{\partial^2}{\partial x_i^2}$. Second form follows from the first one after applying the integration by parts and assuming that the random field $V({\bf x})$
vanishes at infinity. In such an example the role of the operator $\hat{A}$ is obviously
 played by the second-order differential operator $\hat{A}=m^2 -\kappa^2\,\Delta$. Such operators are called "local" as their action on any function
 involves only values of that function and its derivatives in the same point of the space. Knowing $\hat{A}$ explicitly allows one to find
 the two-point correlation function $\langle V({\bf x})\,V({\bf y})\rangle=f({\bf x},{\bf y})$ as the kernel of the operator inverse to $A$ , hence satisfying the differential equation
 \begin{equation}\label{Gaumass1}
\left[m^2 -\kappa^2\,\Delta\right]\,f({\bf x},{\bf y})=\delta({\bf x}-{\bf y}),
\end{equation}
where the Laplacian is assumed to act on the first argument, and $\delta({\bf x}-{\bf y})=\int e^{i{\bf q}({\bf x}-{\bf y})}
\frac{d^N{\bf q}}{(2\pi)^N}$ stands for the appropriate Dirac delta-function.
By applying the Fourier transform to the equation immediately gives the two-point correlation function as
\begin{equation}\label{Gaumass2}
\langle V({\bf x})\,V({\bf y})\rangle=
\int \frac{e^{i{\bf q}({\bf x}-{\bf y})}}{(m^2+\kappa^2 {\bf q}^2)}\frac{d^N{\bf q}}{(2\pi)^N}
\end{equation}
To calculate the above integral it is convenient to use the identity $(m^2+\kappa^2 {\bf q}^2)^{-1}=\int_0^{\infty}e^{-t(m^2+\kappa^2 {\bf q}^2)}\,dt$
and change the order of integration, which gives
\begin{equation}\label{Gaumass3}
\langle V({\bf x})\,V({\bf y})\rangle=\int_0^{\infty}e^{-tm^2}\,dt
\int e^{i{\bf q}({\bf x}-{\bf y})-\kappa^2 t{\bf q}^2}\frac{d^N{\bf q}}{(2\pi)^N}
\end{equation}
\[=\left(\frac{1}{4\pi\kappa^2}\right)^{N/2}
\int_0^{\infty}e^{-tm^2-\frac{1}{4\kappa^2 t}({\bf x}-{\bf y})^2}\,\frac{dt}{t^{N/2}}=\frac{1}{\left(2\pi\right)^{N/2}}
\frac{m^{N/2-1}}{\kappa^{N/2+1}}\frac{K_{N/2-1}\left(\frac{m}{\kappa}|{\bf x}-{\bf y}|\right)}{|{\bf x}-{\bf y}|^{N/2-1}}
\]
where we have used (\ref{Gauprob}) with $A_{ij}\to2\kappa^2 t\delta_{ij}, {\bf b}\to({\bf x}-{\bf y})$ to evaluate the Gaussian
integral in the first line, and $K_\nu(z)$ is the so-called Macdonald function, see the formula 3.471.9 of \cite{GR}.
In particular, for $N=2$ and  $m\to 0$  we have from the expansion 3.471.9 of \cite{GR}
\begin{equation}\label{Gaumass4}
\langle V({\bf x})\,V({\bf y})\rangle=\frac{1}{2\pi\kappa^2}
K_{0}\left(\frac{m}{\kappa}|{\bf x}-{\bf y}|\right)\approx -\frac{1}{2\pi\kappa^2}\ln{\left[\frac{|{\bf x}-{\bf y}|}{2\kappa/m}\right]}, \quad |{\bf x}-{\bf y}|\ll\frac{\kappa}{m}.
\end{equation}
 We conclude that the limit of {\bf 2D massless} Gaussian field provides us with a random field
 with logarithmic correlations.

 The massless Gaussian field is also known in the modern literature as the Gaussian {\it Free} Field (GFF) and considered to be an object of fundamental importance. It can be defined on any domain ${ \bf D}$ of $N-$dimensional space using the following construction.
 Consider an eigenproblem for the Laplace operator $-\Delta$ acting on functions in ${ \bf D}$, and denote ${\bf e}_j({\bf x}),$
   $j=1,2,\ldots, \infty$ its eigenfunctions corresponding to the Dirichlet boundary conditions ( i.e. vanishing at the boundary $\partial {\bf D}$)  and let $\lambda_j>0$ be the corresponding eigenvalues.
  Then the functions $\tilde{{\bf e}}_j({\bf x})=\frac{1}{\sqrt{\lambda_j}}{\bf e}_j({\bf x})$ form an orthonormal basis of the Hilbert space $\texttt{H}$ with respect to the so-called Dirichlet scalar (or "inner") product
  \begin{equation}\label{Dirichlet}
  \left(f,g\right)=\int_{{\bf D}}\left(\nabla f \cdot \nabla g\right)d^N{\bf x}=-\int_{{\bf D}}\left(f \cdot \Delta g\right)d^N{\bf x}
  \end{equation}
  for functions $f({\bf x})$ on ${ \bf D}$ vanishing at the boundary $\partial {\bf D}$. Introduce now a set $\zeta_j, \,\, j=1,2,\ldots, \infty$
  of standard Gaussian independent, identically distributed real variables with mean zero and unit variance each:
  $\langle \zeta_j\rangle=0,\,  \langle \zeta_j^2\rangle=1$. Then the GFF $V({\bf x})$ on the domain ${\bf x}\in {\bf D}$ is defined as
  the formal sum
  \begin{equation}\label{GFF}
  V({\bf x})=\sum_{j=1}^{\infty}\zeta_j\,\tilde{{\bf e}}_j({\bf x}),
  \end{equation}
from which it immediately follows that it is a Gaussian field with the covariance  given by
\begin{equation}\label{GFFcov}
  \left\langle V({\bf x}_1)V({\bf x}_2)\right\rangle =\sum_{j=1}^{\infty}\,\frac{1}{\lambda_j}{\bf e}_j({\bf x}_1){\bf e}_j({\bf x}_2)=-\left(\Delta^{-1}\right)({\bf x}_1,{\bf x}_2)
  \end{equation}
which is nothing else but the Green function $G({\bf x}_1,{\bf x}_2)$ of the Laplace operator on the domain ${\bf D}$. Note however that mathematically $V({\bf x})$
is rather subtle (e.g. the sum in (\ref{GFF}) does not converge pointwise and fails in general
to be an element of the Hilbert space $\texttt{H}$ ).
Because of this and other subtleties an extra mathematical care is needed to define the object fully rigorously,
see references in \cite{DuSh}. The physicists however work with such an object without further ado, and we finish this section
by two simple but important examples. In the first example we deal with the GFF on a one-dimensional domain, the interval ${\bf D}=[0,1]$.
The Laplacian in one dimension is simply $\Delta=-\frac{d^2}{dx^2}$ and the eigenfunctions/eigenvalues of the Dirichlet problem are given by $e_n(x)=\sqrt{2}\sin{n\pi x}, \lambda_n=\pi^2n^2$ so that the GFF in this particular case is given by a random Fourier series
$V(x)=\sum_{n=1}^{\infty} \zeta_n \frac{\sqrt{2}}{\pi n} \sin{n\pi x}$ (compare with the periodic $1/f$ noise in the end of this Appendix). The corresponding
Green function can be easily found to be given by $G(x_1,x_2)=x_1(1-x_2)$ for $ x_2>x_1$ and $G(x_1,x_2)=x_2(1-x_1)$ for $ x_2<x_1$.
One immediately recognizes that the one-dimensional version of the GFF for such a domain coincides with the version of the
Brownian motion called {\it Brownian bridge}, which is conditioned to return to the origin after a given time.

 Our second example is much more relevant in the context of the present lectures
and deals with GFF defined on the two-dimensional disk: ${\bf D}=|z|<L$ where we use the complex coordinate $z=x+iy$.
The Green function for the Dirichlet problem on such a domain is well known and is given by $G(z_1,z_2)=-\frac{1}{2\pi}\ln{\frac{L|z_1-z_2|}{L^2-z_1z_2}}$.   In particular, for any two points $|z_{1,2}|\ll L$ (i.e. well inside the disk)
the Green function reduces to expression equivalent to the full-plane formula  (\ref{Gaumass4}) which is the basis for
models with logarithmic correlations.

Using the full-plane logarithmic GFF it is easy to construct various one-dimensional Gaussian random processes with logarithmic correlations.
In particular, sampling the values of such GFF along a circle of unit radius with coordinates $z=e^{it}, \, t\in[0,2\pi)$ we get a Gaussian process with the covariance $\langle V(t_1)V(t_2)\rangle= -\frac{1}{2\pi}\ln{|e^{it_1}-e^{it_2}|}$.
Such a process can be shown to be equivalent to a random Fourier series of the form $V(t)=\sum_{n=1}^{\infty}
\frac{1}{\sqrt{n}} \left[v_n e^{i n t}+\overline{v}_n e^{-i n t}\right]$, where $v_n,\overline{v}_n$ are independent, identically distributed {\it complex} Gaussian variables with mean zero and variance $\langle v_n \overline{v}_n\rangle=1$ (compare with the earlier Brownian bridge example). As the mean-square value (the "spectral power") of the coefficient in front of a given Fourier harmonic with index $n$ in this case decays like $1/n$ such signals are known in many applications as $1/f$ noises.

\section{Proof of the identity (\ref{trans})}
 We start with identically rewriting the left-hand side of (\ref{trans}) as
 \begin{equation}\label{transa}
  \int {\cal I}\left(X^TX\right)\,d X
= \lim_{\epsilon\to 0^+} \int {\cal I}(Q) {\cal J}_{\epsilon}(Q) \, dQ\,
\end{equation}
where
\begin{equation}\label{transaa}
\quad {\cal J}_{\epsilon}(Q)=\int e^{-\frac{\epsilon}{2} Tr [X^TX]}\delta\left(Q-X^TX\right)\,dX
\end{equation}
and $\delta(x)$ stands for the appropriate Dirac $\delta-$distribution in the matrix space. As usual $\delta-$function
can be expressed via the Fourier transform $\delta(x)=\int e^{-ifx}\frac{df}{2\pi}$ its matrix analogue
can be defined via the following Fourier representation:
\begin{equation}\label{transb}
\fl  \delta\left(Q-X^TX\right)=\int e^{-\frac{i}{2}Tr[(Q-X^TX)F_n]}\,dF_n, \quad dF_n=\prod_{i}\frac{d[F_n]_{ii}}{4\pi}\prod_{i<j}\frac{d[F_n]_{ij}}{2\pi}
\end{equation}
 with the integration going over $n\times n$ real symmetric matrices: $[F_n]_{ij}=[F_n]_{ji}$. Substituting such a representation
 into the expression for ${\cal J}_{\epsilon}(Q)$  and changing the order of integration over $dF_n$ and $dX$ one may
 notice that the integral over $X$ is essentially a product of $N$ identical Gaussian multivariable integrals (\ref{Gaumult})
 where the role of $A$ is played by the matrix $\epsilon {\bf 1}_n-iF_n$. The integrals are well-defined
 due to $\epsilon>0$. Applying (\ref{Gaumult})
 we arrive at ${\cal J}_{\epsilon}(Q)=(2\pi)^{\frac{Nn}{2}}{\cal J}_{n,N,\epsilon}(Q)$, where
 \begin{equation}\label{J1}
 {\cal J}_{n,N,\epsilon}(Q)=\int e^{-\frac{i}{2}Tr[QF_n]} \frac{1}{[\det\left(\epsilon {\bf 1}_n-iF_n\right)]^{N/2}}\,dF_n,
 \end{equation}
and we have indicated explicitly the dependence on $n$ and $N$ for the sake of future reference.
Notice that the integrand is
invariant with respect to the rotations $F\to \hat{O}\hat{F}\hat{O}^{-1}$
where $O$ are orthogonal matrices satisfying $O^TO=1$.  As $Q$ is real symmetric matrix, it can be brought to the diagonal form
by an orthogonal transformation. Hence the result of the integration
can depend only on the eigenvalues $q_1,q_2,...,q_n$ of $\hat{Q}$. Thus, it is enough to
take $\hat{Q}$ to be diagonal from the very beginning. Now we
separate the first eigenvalue from the rest:
\[
\hat{Q}=\mbox{diag}(q_1,q_2,...,q_n)\equiv \mbox{diag}(q_1,\hat{Q}_{n-1})
\]
 and accordingly decompose the
matrix $F_n$ as
\begin{equation}
F_n=\left(\begin{array}{cc}f_{11} & {\bf f}\\
{\bf f}^T& F_{n-1}\end{array}\right)\quad,\quad dF_n=\frac{df_{11}}{4\pi}\frac{d{\bf f}}{(2\pi)^{n-1}}
dF_{n-1}
\end{equation}
where ${\bf f}=\left(f_{12},f_{13},....,f_{1n}\right)$
is a $n-1$ component vector.

Next step is to use the well-known property of the determinants composed of four blocks:
\[
\det{\left(\epsilon{\bf 1}_n-iF_n\right)}=
\det{\left(\epsilon {\bf 1}_{n-1}-iF_{n-1}\right)}
\left(\epsilon-if_{11}+{\bf f}
\left[\epsilon {\bf 1}_{n-1}-iF_{n-1}\right]^{-1}{\bf f}^T\right)
\]
which gives:
\begin{eqnarray}
{\cal J}_{n,N,\epsilon}(\hat{Q})&=&
\int d\hat{F}_{n-1}e^{-\frac{i}{2}\mbox{Tr}\left(\hat{F}_{n-1}\hat{Q}_{n-1}\right)}
\left[\det{\left(\epsilon {\bf 1}_{n-1}-iF_{n-1}\right)}\right]^{-N/2}\\
\nonumber &\times& \int \frac{d{\bf f}}{(2\pi)^{n-1}}\int_{-\infty}^{\infty}
\frac{df_{11}}{4\pi}e^{-\frac{i}{2}f_{11}q_1}\frac{1}
{\left(\epsilon-if_{11}+{\bf f}
\left[\epsilon {\bf 1}_{n-1}-iF_{n-1}\right]^{-1}{\bf f}^T\right)^{N/2}}
\end{eqnarray}
The last integral over $f_{11}$ can be explicitly evaluated by using the formula 3.382.7 of \cite{GR}:
\begin{equation}\label{idcauchy}
\int e^{-ifp}\frac{1}{(\beta-if)^{\nu}}\frac{df}{2\pi}=\frac{p^{\nu-1}}{\Gamma{(\nu)}}e^{-\beta p}\theta(p), \quad Re(\nu,\beta)>0
 \end{equation}
where $\Gamma(\nu)$ is the Euler Gamma-function, and $\theta(x)=1$ for $x>0$ and zero otherwise.
Taking into account $\epsilon>0$, the result of the
 integration over $f_{11}$ gives
\begin{equation}
\frac{1}{2\Gamma(N/2)}\theta(q_1)\left(\frac{q_1}{2}\right)^{N/2-1}
\exp\left\{-\frac{1}{2}q_1\left(\epsilon+{\bf f}
\left[\epsilon {\bf 1}_{n-1}-iF_{n-1}\right]^{-1}{\bf f}^T\right)\right\}\,.
\end{equation}
 Now the integration over the vector $d{\bf f}$ becomes the standard Gaussian and can be performed using (\ref{Gaumult})
 yielding the factor:
\[
\left(\frac{1}{2\pi q_1}\right)^{\frac{n-1}{2}}
\mbox{det}^{1/2}\left(\epsilon {\bf 1}_{n-1}-iF_{n-1}\right)
\]
Collecting all the factors we arrive at the recursive relation
\begin{equation}
{\cal J}_{n,N,\epsilon}(\hat{Q})=\frac{\pi^{-\frac{n-1}{2}}}{2^n\Gamma(N/2)}\left(\frac{q_1}{2}\right)^{\frac{N-n-1}{2}}
 \theta(q_1)e^{-\frac{1}{2}\epsilon q_1}{\cal J}_{n-1,N-1,\epsilon}(\hat{Q}_{n-1})
\end{equation}
 This relation can be iterated further, and assuming $N>n$ we arrive at the last step to (\ref{idcauchy}) which gives
\begin{equation}\label{idcauchy1}
{\cal J}_{1,N-n+1,\epsilon}(q_{n})=\frac{1}{2\Gamma(\frac{N-n+1}{2})}\left(\frac{q_n}{2}\right)^{\frac{N-n-1}{2}}
 \theta(q_n)e^{-\frac{1}{2}\epsilon q_n}
\end{equation}
and serves as an "initial condition" for our iteration scheme. This immediately
yields the result:
\begin{equation}
\fl {\cal J}_{n,N,\epsilon}(Q)=\frac{1}{2^{\frac{Nn}{2}}\pi^{\frac{n(n-1)}{4}}}
\frac{1}{\prod_{j=0}^{n-1}\Gamma\left(\frac{N-j}{2}\right)}
\, \mbox{det}^{\frac{N-n-1}{2}}
\left[Q\right]
e^{-\frac{1}{2}\epsilon\mbox{Tr}\,Q}\,\, \prod_{j=1}^n\theta(q_j)
\end{equation}
for $N\ge n+1$. As ${\cal J}_{\epsilon}(Q)=(2\pi)^{\frac{Nn}{2}}{\cal J}_{n,N,\epsilon}(Q)$, in the limit $\epsilon\to 0$
 the above relation yields precisely the required identity (\ref{trans}).

 \section{ Parisi matrix, its eigenvalues and evaluation of traces
in the replica limit.}\label{AppendixB}

We start with describing the well known structure of the $n\times
n$ matrix $Q$ in the Parisi parametrisation, see Fig.12a. At the beginning we
set $n$ diagonal entries $q_{\alpha\alpha}$ all to the same value
$q_{\alpha\alpha}=0$. This value will be maintained at every but
last step of the recursion. The off-diagonal part of the matrix
$Q$ in the Parisi scheme is built recursively as follows. At the
first step we single out from the $n\times n$ matrix $Q$ the chain
consisting of $n/m_{1}$ blocks of the size $m_1\le n$, each
situated on the main diagonal. All off-diagonal entries
$q_{\alpha\beta},\,\alpha\ne \beta$ inside those blocks are filled
in with the same value $q_{\alpha\beta}=q_1\le 0$, whereas all the
remaining $n^2(1-1/m_1)$ entries of the matrix $Q$ are set to the
value $0<q_0\le q_1$. The latter entries remain from now on intact
to the end of the procedure, whereas some entries inside the
diagonal $m_1\times m_1$ blocks will be subject to a further
modification. At the next step of iteration in each of those
diagonal blocks of the size $m_1$ we single out the chain of
$m_2/m_1$ smaller blocks of the size $m_2\le m_1$, each situated
on the main diagonal. All off-diagonal entries
$q_{\alpha\beta},\,\alpha\ne \beta$ inside those sub-blocks are
filled in with the same value $q_{\alpha\beta}=q_2\ge q_1$,
whereas all the remaining entries of the matrix $Q$ hold their old
values. At the next step only some entries inside diagonal blocks
of the size $m_2$ will be modified., etc. Iterating this procedure
step by step one obtains after $k$ steps a hierarchically built
structure characterized by the sequence of integers
\begin{equation}\label{parisiseq}
n=m_0\ge m_1\ge m_2\ge \ldots\ge m_k\ge m_{k+1}=1
\end{equation}
and the values placed in the diagonal blocks of the $Q$ matrix
satisfying:
\begin{equation}\label{parisiseq1}
0<q_0\le q_1\le q_2\le \ldots\le q_k
\end{equation}
Finally, we complete the procedure by filling in the $n$ diagonal
entries $q_{\alpha\alpha}$ of the matrix $Q$ with one and the same
value $q_{\alpha\alpha}=q_d\ge q_k$.

For the subsequent analysis we need the eigenvalues of the Parisi
matrix $Q$. Those can be found easily together with the
corresponding eigenvectors built according to a recursive
procedure which uses the sequence Eq.(\ref{parisiseq}). It is
convenient to visualize eigenvectors as being "strings" of $n$
boxes numbered from $1$ to $n$, with $l^{th}$ component being a
content of the box number $l$.

At the first step $i=1$ we choose the eigenvector to have all $n$
boxes filled with the same content equal to unity. The
corresponding eigenvalue is non-degenerate and equal to
\begin{equation}\label{parisieigen1}
\lambda_1=q_d+q_k(m_k-1)+q_{k-1}(m_{k-1}-m_k)+
\ldots+q_1(m_1-m_2)+q_0(m_0-m_1)
\end{equation}
Now, at the subsequent steps $i=2,3,\ldots,k+2$ one builds
eigenvectors by the following procedure. The string of $n$ boxes
of an eigenvector belonging to $i^{th}$ family are subdivided into
$n/m_{i-1}$ substrings of the length $m_{i-1}$, and numbered
accordingly by the index $j=1,2,\ldots, n/m_{i-1}$. All $m_{i-1}$
boxes of the first substring $j=1$ are filled invariably with all
components equal to $1$. Next we fill $m_{i-1}$ boxes in one (and
only one) of the remaining $\frac{n}{m_{i-1}}-1$ substrings with
all components equal to $-1$. In doing so we however impose a
constraint that the substrings with the indices $j$ given by
$j=1+l\frac{m_{i-2}}{m_{i-1}}$ should be excluded from the
procedure, with $l$ being any integer satisfying $1\le l\le
\frac{n}{m_{i-2}}-1$. After the choice of a particular substring
is made, we fill all $n-2m_{i-1}$ boxes of the remaining
substrings with identically zero components. It is easy to see
that all $d_i=n/m_{i-1}-n/m_{i-2}$ different eigenvectors of
$i^{th}$ family built in such a way correspond to one and the same
$d_i-$degenerate eigenvalue
\begin{equation}\label{parisieigeni} \lambda_i=q_d+q_k(m_k-1)+q_{k-1}(m_{k-1}-m_k)+
\ldots+q_{i-1}(m_{i-1}-m_i)-q_{i-2}(m_{i-1})
\end{equation}
In this way we find all $n$ possible eigenvalues, the last being
equal to
\begin{equation}\label{parisieigenk} \lambda_{k+2}=q_d-q_k\,
m_{k+1}\equiv q_d-q_k.
\end{equation}
The completeness of the procedure follows from the fact that sum
of all the degeneracies $d_i$ is equal to
\[
1+\left(\frac{n}{m_1}-1\right)+\left(\frac{n}{m_2}-\frac{n}{m_1}\right)+\ldots+
\left(\frac{n}{m_{k+1}}-\frac{n}{m_k}\right)=n
\]
Note that all the found eigenvalues are positive due to
inequalities Eq.(\ref{parisiseq1}) between various $q_i$, which is
required by the positive definiteness of the matrix $Q$. Note also
that all eigenvectors built in this way are obviously linearly
independent, although the eigenvectors belonging to the same
family are not orthogonal. The latter fact however does not have
any bearing for our considerations.

To facilitate the subsequent treatment it is convenient to
introduce the following (generalized) function of the variable
$q$, see Fig.12b:
\begin{equation}\label{xstep}
x(q)=n+\sum_{l=0}^k (m_{l+1}-m_l)\,\theta(q-q_l)
\end{equation}
where we use the notation $\theta(z)$ for the Heaviside step
function: $\theta(z)=1$ for  $z>0$ and zero otherwise. In view of
the inequalities Eq.(\ref{parisiseq},\ref{parisiseq1}) the
function $x(q)$ is piecewise-constant non-increasing, and changes
between $n$ and $1$ as follows:
\begin{equation}\label{xstep1}
\fl x(q<q_0)=m_0\equiv
n,\,\,x(q_0<q<q_1)=m_1,\,\ldots,\,x(q_{k-1}<q<q_k)=m_k,\,x(q>q_k)=m_{k+1}\equiv
1
\end{equation}
Comparison of this form with Eq.(\ref{xstep}) makes evident the
validity of a useful inversion formula:
\begin{equation}\label{xstep2}
\frac{1}{x(q)}=\frac{1}{n}+\sum_{l=0}^k
\left(\frac{1}{m_{l+1}}-\frac{1}{m_l}\right)\,\theta(q-q_l)
\end{equation}
which will be exploited by us shortly.

 As observed by Crisanti and Sommers\cite{sph1} one
can represent the eigenvalues Eq.(\ref{parisieigeni}) of the
Parisi matrix in a compact form via the following remarkable
identities:
\begin{equation}\label{criso}
\lambda_{1}=\int_{0}^{q_d}x(q)\,dq=nq_0+\int_{q_0}^{q_d}x(q)\,dq,\quad\lambda_{i+2}=\int_{q_i}^{q_d}x(q)\,dq,\quad
i=0,1,\ldots,k
\end{equation}
As a consequence, these relations imply for any analytic function
$g(x)$ the identity
\begin{equation}\label{criso1}
\fl \frac{1}{n}Tr\left[
g(Q)\right]=\frac{1}{n}\sum_{i=1}^{k+2}g(\lambda_{i})\,d_i
=\frac{1}{n}g\left(nq_0+\int_{q_0}^{q_d}x(q)\,dq\right)+
\sum_{l=0}^{k}\left(\frac{1}{m_{l+1}}-\frac{1}{m_{l}}\right)g\left(\int_{q_l}^{q_d}x(q)\,dq\right)
\end{equation}
Next one observes that taking the derivative of the generalized
function from Eq.(\ref{xstep2}) produces
\begin{equation}\label{xstep3}
\frac{d}{dq}\left[\frac{1}{x(q)}\right]=\sum_{l=0}^k
\left(\frac{1}{m_{l+1}}-\frac{1}{m_l}\right)\,\delta(q-q_l).
\end{equation}
This fact allows one to rewrite the sum in Eq.(\ref{criso1}) in
terms of an integral, yielding
\[
\frac{1}{n}Tr\left[ g(Q)\right]
=\frac{1}{n}g\left(nq_0+\int_{q_0}^{q_d}x(q)\,dq\right)+
\int_{q_0-0}^{q_k+0}g\left(\int_{q}^{q_d}x(\tilde{q})\,d\tilde{q}\right)\,\frac{d}{dq}\left[\frac{1}{x(q)}\right]\,dq,
\]
where the short-hand notation $q\pm 0$ designates the limit from
below/above. Further performing integration by parts, and using
$x(q>q_k)=1,\,x(q<q_0)=n$, we finally arrive at
\begin{equation}\label{criso11}
\fl \frac{1}{n}Tr\left[ g(Q)\right]
=\frac{1}{n}\left[g\left(nq_0+\int_{q_0}^{q_d}x(q)\,dq\right)-g\left(\int_{q_0}^{q_d}x(q)\,dq\right)\right]
+\int_{q_0}^{q_k}g'\left(\int_{q}^{q_d}x(\tilde{q})\,d\tilde{q}\right)dq+g(q_d-q_k).
\end{equation}
We are actually interested in the replica limit $n\to 0$.
According to the Parisi prescription in such a limit the
inequality Eq.(\ref{parisiseq}) should be reversed:
\begin{equation}\label{parisiseq2a}
n=0\le m_1\le m_2\le \ldots\le m_k\le m_{k+1}=1
\end{equation}
and the function $x(q)$ is now transformed to a non-decreasing
function of the variable $q$ in the interval $q_0\le q \le q_k$,
and satisfying outside that interval the following properties
\begin{equation}\label{outsidea}
x(q<q_0)=0,\quad \mbox {and}\quad  x(q>q_k)=1.
\end{equation}
In general,such a function also depends on the increasing sequence
of $k$ parameters $m_l$ described in Eq.(\ref{parisiseq2}) .

The form of Eq.(\ref{criso11}) makes it easy to perform the limit
$n\to 0$ explicitly, and to obtain after exploitation of
Eq.(\ref{outside}) an important identity Eq.(\ref{traces}) helping
to evaluate the traces in the replica limit. Finally, let us
mention the existence of an efficient method of the "replica
Fourier transform" allowing one to diagonalise (and otherwise
work) with much more general types of hierarchical matrices, see
\cite{Dedominicis} for more details.

 \section{Stability analysis of the saddle-point solution}

Our starting point is the functional $\Phi_n(Q)$ from (\ref{repham1}) whose extrema we look for in the space of positive definite
matrices $Q$ constrained to have the diagonal entries $q_{aa}=R^2$. The independent variables are all off-diagonal entries $q_{(ab)}$
where $(ab)$ stands for $n(n-1)/2$ "ordered" pairs with $a<b$, and the stationary values are found from the equations (\ref{sp2}).
The stability matrix in this space is given by $A_{(ab),(cd)}=\frac{\partial^2}{\partial q_{(ab)}\partial q_{(cd)}}\Phi_n(Q)$ which should be evaluated
at the saddle-point solution.  In a general situation we should distinguish three types of entries of that matrix: the diagonal entries
\begin{equation}\label{st1}
A_{(ab),(ab)}=\left[\left(Q^{-1}\right)_{aa}\left(Q^{-1}\right)_{bb}+\left(Q^{-1}\right)^2_{ab}\right]-\gamma\frac{1}{(R^2-q_{ab}+1)^2}\,,
\end{equation}
the entries for the case when the ordered pairs $(ab)$ and $(cd)$ share one common replica, that is
\begin{equation}\label{st2}
A_{(ab),(ac)}=\left[\left(Q^{-1}\right)_{aa}\left(Q^{-1}\right)_{bc}+\left(Q^{-1}\right)_{bc}\left(Q^{-1}\right)_{ab}\right], \,\, b<c
\end{equation}
and a similar expression for $A_{(ab),(cb)}\,, a<c$, and finally the entries for the ordered pairs $(ab)$ and $(cd)$ which do not share any common replica:
\begin{equation}\label{st3}
A_{(ab),(cd)}=\left[\left(Q^{-1}\right)_{ac}\left(Q^{-1}\right)_{bd}+\left(Q^{-1}\right)_{ad}\left(Q^{-1}\right)_{bc}\right], \,\,
\end{equation}
If we are interested in investigating stability of the replica-symmetric solution, we should substitute to the above equations
$q_{ab}=q_0, \forall a\ne b$ as well as $\left(Q^{-1}\right)_{aa}=p_d, \forall a$ and $\left(Q^{-1}\right)_{ab}=p_0, \forall a<b$, with $p_d$ and $p_0$ taken from (\ref{invsym1},\ref{invsym2}). This gives for the entries of the stability matrix
\begin{equation}\label{st4}
 \fl A_{(ab),(ab)}\equiv A_1=p_d^2+p_0^2-\gamma\frac{1}{(R^2-q_{0}+1)^2}\,,
\quad A_{(ab),(ac)}\equiv A_2=p_0p_d+p_0^2
\end{equation}
and $\quad A_{(ab),(cd)}\equiv A_3=2p_0^2$. As discovered by De Almeida and Thouless \cite{AT} the eigenvalues/eigenvectors
of such $n(n-1)/2\times n(n-1)/2$ matrix can be found explicitly. There are three families of eigenvectors. The first family consists
of a single "replica-symmetric" eigenvector ${\bf e}_{1}$ with all components $\left[{\bf e}_{1}\right]_{(ab)}=1$.
The corresponding eigenvalue is equal to the sum of all entries in one row of $A$ that is
$\lambda_1=A_1+2(n-2)A_2+\frac{(n-2)(n-3)}{2}A_3$. Next family consists of $d_2=n-1$ eigenvectors ${\bf e}_{2}^{c}, \, c=1,\dots n-1 $ with one
replica index $c$ singled out. For example, suppose that $c=1$, then  $\left[{\bf e}^1_{2}\right]_{(ab)}=\frac{n-2}{2}$ if $a=1$ or $b=1$,
and  $\left[{\bf e}^1_{2}\right]_{(ab)}=-1$ otherwise (note that such eigenvector is orthogonal to ${\bf e}_{1}$). The corresponding eigenvalue shared by all the eigenvectors in the family is $\lambda_2=A_1+(n-4)A_2-(n-3)A_3$. Finally , third family consists of $d_3=\frac{n(n-3)}{2}$
eigenvectors ${\bf e}^{(cd)}_{3}$ with an ordered pair of replica indices $c<d$ singled out. For example, if $(cd)=(12)$ then components of the corresponding eigenvector are $\left[{\bf e}^{(12)}_{3}\right]_{(12)}=\xi$ , $\left[{\bf e}^{(11)}_{3}\right]_{(ab)}=\psi$ if $a=1,2$ or $b=1,2$,
and otherwise  $\left[{\bf e}^{(11)}_{3}\right]_{(ab)}=\rho$ where the values of $\xi,\psi,\rho$ should be chosen to make ${\bf e}^{(11)}_{3}$
orthogonal to ${\bf e}^{(1)}_{2}$ and ${\bf e}_{1}$. The eigenvalue shared by the third family turns out to be given by an $n-$independent expression $\lambda_3=A_1-2A_2+A_3$. Since $1+d_2+d_3=n(n-1)/2$ no more eigenvalues are possible.

It is well known in general (and can be easily checked for our model)
that it is third family which gives rise to "dangerous" fluctuations breaking down the replica symmetry of the saddle-point solution
in the limit $n\to 0$ below some critical temperature $T_c$ at which $\lambda_3$ vanishes.  Substituting the expressions (\ref{st4}) into $\lambda_3$ and using the relation (\ref{invsym}) we find that the condition $\lambda_3=0$ is equivalent to
\begin{equation}\label{st5}
\frac{1}{R^2-q_0}=\frac{\gamma^{1/2}}{R^2-q_0+1}\,.
\end{equation}
Finally, using for the combination $R^2-q_0\equiv d_0$ the equation (\ref{sp1sym1}) we find after simple algebra that the critical value of
the parameter $\sqrt{\gamma}=\beta g$ is given by $\sqrt{\gamma}_c=\frac{R^2+1}{R^2-1}$ as was quoted in the text.

 Let us now turn to the stability issue for the one-step RSB solution which we claimed to be the correct choice
below  the critical temperature. According to Fig.12a the one-step solution is characterized by the matrices $Q$ with all diagonal entries still
equal to $R^2$, and two different values of the off-diagonal entries $q_1>q_0$. The size of blocks containing $q_1$ is equal to $m$.
Such more complicated structure of $Q$ generates more types of different
elements in the stability matrix $A$, and although its eigenvalues/eigenvectors can be still successfully found \cite{FS}, actual
analysis becomes long. Referring the interested reader to Appendix B3 of \cite{FS} for  a detailed exposition, we give below a
very brief summary of the the outcome of the procedure.
Actually the stability analysis of \cite{FS} was performed for models with general random potential characterized
by the covariance $\left\langle V\left({\bf x}_1\right) \, V\left({\bf
x}_2\right)\right\rangle=\,Nf\left(({\bf x}_1-{\bf
x}_2)^2/2N\right)$.
It was found that the matrix $A$ in that general case has nine different eigenvector families, and the stability
is controlled by two of them with eigenvalues given by
\begin{equation}
\fl  \Lambda^*_0= \left[{1\over R^2-q_1 +m(q_1-q_0)} - {1\over T} \sqrt{f''(R^2-q_0)}\right] \left[{1\over R^2-q_1 +m(q_1-q_0)} + {1\over T} \sqrt{f''(R^2-q_0)}\right]
\label{lambda0}
\end{equation}
and
\begin{equation}
\Lambda^*_K= \left[{1\over R^2-q_1} - {1\over T} \sqrt{f''(R^2-q_1)}\right]\left[{1\over R^2-q_1} + {1\over T} \sqrt{f''(R^2-q_1)}\right]\,,
\label{lambdaK}
\end{equation}
assuming $f''(x)>0$. If both of the above eigenvalues are non-negative, and the second derivative $f''(x)$ is monotonically decreasing with $x$ then
 all the remaining seven eigenvalues are strictly positive and the system is stable.

It is easy to understand that the logarithmic case (\ref{2})
considered in this paper is recovered, after all due rescalings in the limit $N\gg 1$, by the choice $f(x)=-g^2\ln{(x+1)}$
so that $\frac{1}{T}\sqrt{f''(x)}=\frac{\sqrt{\gamma}}{x+1}$. Using now the equilibrium values for the parameters $q_1,q_0,m$ given in (\ref{1step})
one finds after a straightforward algebra that the first brackets in (\ref{lambda0}),(\ref{lambdaK}) identically vanish leaving us with $\Lambda^*_0=\Lambda^*_K=0$
{\it everywhere} in the low-temperature phase. This implies indeed that the corresponding one-step RSB solution is {\it marginally} stable.

\end{document}